\documentclass[12pt,preprint]{aastex}

\slugcomment{To appear in ApJ}

\shorttitle{The Disk Fraction of wTTs }
\shortauthors{Cieza et al.}

\begin{document}

\title{The \emph{Spitzer} c2d Survey of Weak-line T Tauri Stars II: New Constraints on the Timescale for Planet Building}
\author{Lucas Cieza\altaffilmark{1},
Deborah L. Padgett\altaffilmark{2},
Karl R. Stapelfeldt\altaffilmark{3}, 
Jean-Charles Augereau\altaffilmark{4},
Paul Harvey\altaffilmark{1}, 
Neal J. Evans, II\altaffilmark{1},
Bruno Mer\'{\i}n \altaffilmark{5},
David Koerner\altaffilmark{6}, 
Anneila Sargent\altaffilmark{7},
Ewine F. van Dishoeck\altaffilmark{5},
Lori Allen\altaffilmark{8}, 
Geoffrey  Blake\altaffilmark{7}, 
Timothy Brooke\altaffilmark{7}, 
Nicholas Chapman\altaffilmark{9},  
Tracy Huard\altaffilmark{8},
Shih-Ping Lai\altaffilmark{9}, 
Lee Mundy\altaffilmark{9}, 
Philip C. Myers\altaffilmark{8}, 
William Spiesman\altaffilmark{1}, 
Zahed Wahhaj\altaffilmark{6}}

\altaffiltext{1}{Astronomy Department, University of Texas, 1 University
Station C1400, Austin, TX 78712} 
\altaffiltext{2}{{\it Spitzer} Science Center, MC220-6, California Institute of Technology, Pasadena, CA 91125}
\altaffiltext{3}{Jet Propulsion Laboratory, MS 183-900, 4800 Oak Grove Drive,
Pasadena, CA 91109}
\altaffiltext{4}{Laboratoire d'Astrophysique de Grenoble, BP 53, 38041 Grenoble Cedex 9}
\altaffiltext{5}{Leiden Observatory, Postbus 9513, 2300 R.A. Leiden, Netherlands}
\altaffiltext{6}{Northern Arizona University, Department of Physics and Astronomy,
Box 6010, Flagstaff, AZ 86011}
\altaffiltext{7}{California Institute of Technology, Pasadena, CA 91125}
\altaffiltext{8}{Smithsonian Astrophysical Observatory, 60 Garden Street,
MS42, Cambridge, MA 02138}
\altaffiltext{9}{Astronomy Department, University of Maryland, College Park, MD
20742}
         
\begin{abstract}

One of the central goals of the \emph{Spitzer} Legacy Project ``From Molecular Cores to Planet-forming Disks'' (c2d) is to 
determine the frequency of remnant circumstellar disks around weak-line T Tauri stars (wTTs) and to study the properties 
and evolutionary status of these disks. Here we present a census of disks for a sample of over 230 spectroscopically  
identified wTTs located in the c2d  IRAC (3.6, 4.5, 4.8, and 8.0 $\mu$m) and MIPS (24 $\mu$m) maps of the Ophiuchus, Lupus, and 
Perseus Molecular Clouds. We find that $\sim$20$\%$ of the wTTs in a magnitude limited subsample  have noticeable IR-excesses 
at IRAC wavelengths indicating the presence of a circumstellar disk. The disk frequencies we find in these 3 regions are $\sim$3-6 times 
larger than that recently found for a sample of 83 relatively isolated wTTs located, for the most part, outside the highest extinction regions 
covered by the c2d IRAC and MIPS maps. This discrepancy in the disk fraction of these two different groups of wTTs (on cloud vs. 
off-cloud targets) supports the idea that samples of wTTs distributed \emph{around} molecular clouds
(nominally 1-10 Myrs old)  represent a somewhat older population of stars. The disk fractions we find are more consistent with those obtained
in recent \emph{Spitzer} studies of wTTs in young clusters such as IC 348 and Tr 37.  
From their location in the H-R diagram, we find that, in our sample, the wTTs with excesses  are among the younger part of the 
age distribution. Still, up to $\sim$50$\%$ of the apparently  youngest stars in the sample show no 
evidence of IR excess, suggesting that the circumstellar disks of a  sizable fraction of pre-main-sequence stars dissipate in a 
timescale of $\sim$1 Myr. We also find that none of the stars in our sample apparently older than $\sim$10 Myrs have detectable 
circumstellar disks at wavelengths $\leq$ 24 $\mu$m. Our result on the survival time of primordial disks is very similar to those 
obtained by studies based on K-band identified inner disks (r $<$ 0.1 AU). 
Also, we find that the wTTs disks in our sample exhibit a wide range of properties (SED morphology, inner radius, 
L$_{DISK}$/L$_{*}$, etc) which bridge the gaps observed between the cTTs and the debris disk regimes. This strongly 
suggests that wTTs disks are in fact the link between the massive primordial disks found around cTTs and the debris disks 
observed around main-sequence stars.

\end{abstract}

\section{Introduction}

Over the last couple of decades, it has been clearly established that circumstellar disks are an integral part of the star 
formation process. Even though there is currently no direct evidence that planets actually grow from circumstellar material, it has 
become increasingly clear that disks are potential birthplaces of planets because their masses, sizes, 
and compositions are consistent with those of the assumed pre-planetary solar nebulae (Hillenbrand 2003). More recently, the
discovery of exo-planets orbiting nearby main sequence stars has confirmed that the formation of planets is a common 
process and not a rare phenomenon exclusive to the Solar System. 

Direct detection of forming planets is well beyond our current capabilities and observing molecular hydrogen, 
which largely dominates the mass of primordial disks, is particularly challenging (Thi et al. 2001, Richter et al. 2002).
However, the thermal emission from circumstellar dust is much easier to detect and study. For this reason, 
the study of  the evolution of circumstellar dust has been a natural first step toward providing observational constraints 
on planet formation theories. 
 
Strom et al. (1989)  studied a sample of 83 classical  T Tauri stars (cTTs) and weak-line T Tauri stars (wTTs) located in the 
Taurus-Auriga star-forming region in order to  determine the fraction of objects with K-band (2.2 $\mu$m) and IRAS excesses 
indicating the presence of a circumstellar disk. WTTs are low-mass pre-main sequence stars that occupy the same region of 
the H--R diagram as cTTs but do not show clear evidence of accretion. The distinction between the two is usually made 
based on the $H\alpha$ equivalent width (EW). The $H_\alpha$ EW of cTTs is $>$ 10 $\AA$,  while the $H\alpha$ EW of wTTs 
is $<$ 10 $\AA$. Since there is a very strong correlation between spectroscopic signatures of gas accretion and the presence 
of near-IR excess (Hartigan et al. 1995), most CTTS show near-IR excess while most wTTs lack such an excess.
Strom et al. (1989) found that 60$\%$ of their stars younger than 3 Myrs showed a K-band excess, indicating the presence 
of a circumstellar disk, while only 10$\%$ of the stars older than 10 Myrs years did so. Based on  these numbers, 
they estimated a disk dissipation  timescale of $<$ 3-10 Myrs and claimed that their result  was, at the time, 
``the best astrophysical constraint on the time available for planet building''. It has been argued that
individual star-forming regions such as Taurus lack the intrinsic age spread necessary to investigate the 
dissipation timescale of circumstellar disks from individually derived ages (Hartmann 2001). However, similar disk 
lifetimes studies  based on the disk frequency of clusters with different mean ages and extending to the 3.4 $\mu$m L-band
(e.g. Haisch et al. 2001) have lead to results similar to those presented by Strom and collaborators 
(see Hillenbrand (2006) for a review on the frequency of near-IR excesses based on a sample of $\sim$3000 PMS). 
The K-band excess, when used as a disk indicator,  is only sensitive to dust in the innermost part of the disk; therefore, 
 K-band studies only constrain the dissipation timescale of a region of the disks that is much closer to the star 
than the locations corresponding to the orbits of any of the planets in the Solar System (Mercury's semi-major axis is 0.39 AU). 
The dissipation timescale of the dust in the planet-forming regions might or might not be the same. Since the dynamical timescale 
is shorter and the surface density is higher closer to the central star, circumstellar disks are expected to evolve from the inside 
out. Most wTTs, which by definition present little or no evidence for accretion, also show little or no near-IR excess (Hartigan et al. 1995). 
However, even after the inner accretion disk has dissipated, it is entirely possible that wTTs to still have enough material at 
larger radii to form terrestrial and giant planets. In fact, millimeter-wave observations show that at least 10 $\%$ of wTTs have 
disks with estimated masses in the $10^{-1}$--$10^{-3}$ $M_{\odot}$ range (Osterloh $\&$ Beckwith 1995, Andrews and Williams 2005).

While the existence of planets around a significant fraction of all MS stars has been verified (e.g., Marcy $\&$ Butler 1998), the 
fundamentals of the planet formation process still remain open questions, especially for giant planets. There are currently three 
main theories for the formation of giant planets: core accretion, gravitational instability, and hybrid models that combine aspects 
of both theories. See Lissauer $\&$ Stevenson (2007) and Durisen et al. (2007) for two recent reviews of the core 
accretion and  gravitational instability models and a discussion of the many upstanding questions. 
   
Although it is unlikely that an observational estimate of the disk's  dissipation timescale by itself can distinguish between the 
competing theoretical models mentioned above, estimates of the dissipation timescale of the planet forming regions can impose 
valuable constraints on current theoretical models. In order to probe the planet-forming regions of disks around pre-main sequence (PMS) 
stars, mid- and far-IR observations are required. Unfortunately, these spectral regions are not easily observable from the ground, 
and past space IR telescopes such as IRAS and ISO were only sensitive enough to detect very bright optically thick disks
in low-mass stars at the distance of nearest star-forming regions. These instruments lacked the sensitivity needed to detect the 
modest IR-excesses expected for optically thin disks and faint optically thick disks. Thus, \emph{Spitzer}'s sensitivity is required 
to establish whether most wTTs have optically thin disks, disks with inner holes, which are too cold to be detected in the near-IR from 
the ground, disks too faint to be detected by IRAS and ISO, or no disks at all. One of the main goals of the \emph{Spitzer}  Legacy Project 
``From Molecular Cores to Planet-forming Disks'' (c2d; Evans et al. 2003) is to determine whether or not most wTTs have circumstellar 
disks and to characterize their properties and evolutionary status. Preliminary results from the c2d Legacy Project (Padgett et al. 
2006, P06 hereafter) showed that disks are rare ($\sim$6$\%$) among the population of wTTs distributed \emph{around} nearby molecular 
clouds.  However, other recent \emph{Spitzer} studies have reported significantly larger disk fractions ($\sim$30$\%$) among wTTs in 
young clusters such as IC 348 and Tr 37 (Lada et al. 2006; Sicilia-Aguilar et al 2006). Here we study a sample of over 230 spectroscopically 
identified wTTs located in the c2d  IRAC and MIPS maps of the Lupus, Ophiuchus, and Perseus Molecular Clouds in order to investigate the 
frequency of circumstellar disks as a function of stellar age. In Section 2, we describe the c2d survey of molecular clouds and our sample 
of wTTs.  In Section 3, we identify IR excesses and investigate the properties of their disks. In Section 4, we compare our results to 
previous \emph{Spitzer} results and discuss the evolutionary status of wTTs disks. Also in Section 4, we derive the ages of the wTTs in 
our sample from their location in the H-R diagram. We investigate the disk frequency as a function of stellar age and use our results to 
impose constraints on the timescale for planet building. Finally, our conclusions are summarized in Section 5.

\section{Observations}\label{observations_01}
\subsection{C2D large molecular clouds and GTO  observations}
 
As part of the c2d Legacy  Project, \emph{Spitzer}  has mapped 13.0 sq. deg. of 3 nearby star-forming 
regions, Perseus, Ophiuchus, and  Lupus, with IRAC (3.6, 4.5, 5.8, and 8.0 $\mu$m) and 22.1 sq. deg. 
with MIPS (24 $\mu$m). MIPS 70 and 160 $\mu$m observations were also taken, but due to sensitivity 
considerations, we do not include these observations in  most of our analysis. The IRAC maps consist of two dithers 
of 10.4 sec observations, each obtained at two epochs (41.6 sec total) separated by several hours. The second epoch 
observations were taken in the High Dynamic Range mode, which includes 0.4 sec observations before the 10.4 sec exposures, 
allowing photometry of both bright and faint stars at the same time. MIPS observations were taken with the fast 
scan mode, also in  two different epochs of 15 sec exposures each. See Jorgensen et al. (2006) and Young et al. (2005) 
for a detailed description of the observing strategy used for the c2d IRAC and MIPS survey of nearby molecular clouds.
      
    In addition to the data from the c2d legacy project, we use observations of IC\,348 taken as part of 
the IRAC and MIPS Guaranteed Time Observer (GTO) programs (Program ID 36 and 58, respectively). 
The IRAC GTO  observations cover a 15$\arcmin$\,x\,15$\arcmin$ field of view centered in the cluster 
and consists of two pairs of 8 dithers of 96.8 sec exposures for the 3.6, 4.5, and 5.8\,\micron\ observations 
(e.g. 1549 sec exposures per pixel). The 8.0\,\micron\ observations consist of four pairs of 8 dithers of 46.8 sec 
exposures \footnote{The longest integration time of the 8.0 $\mu$m  array, nominally 100 secs, consists of 2 exposures 
of 46.8 sec each (See IRAC handbook, http://ssc.spitzer.caltech.edu/irac/dh/).}. The MIPS 24 $\mu$m GTO observations of 
IC 348 were taken in the medium scan mode resulting in an average exposure time of 80 seconds per pixel. For consistency, 
we processed the Basic Calibrated Data from the GTO programs and produced point source catalogs using the c2d pipeline 
(Evans et al. 2005). The c2d pipeline uses the the c2d mosaicking/source extraction software c2dphot (Harvey et al. 2004), 
which is based on the mosaicking program APEX developed by the \emph{Spitzer} Science Center, and the source 
extractor Dophot (Schechter et al. 1993).

Flux uncertainties in c2dphot are calculated in a standard way from a numerical estimate of the Hessian matrix 
(Press et al. 1992; Silvia 1996). This procedure for estimating uncertainties, although statistically correct, appears to 
underestimate the uncertainty as measured by the repeatability of flux measurements of the same objects at different epochs. 
For bright sources, there appears to be a random error floor to the best uncertainty possible with our observing techniques 
of 0.05 mag for the IRAC bands and 0.09 mag for the MIPS bands.
The absolute calibration uncertainties are not included in our uncertainties. They are 5 and 10$\%$ for IRAC and MIPS, 
respectively (see data handbook for the instruments). 
As in most \emph{Spitzer} surveys, the  intrinsic sensitivity of the c2d observations is not uniform 
across the clouds due to variations in the total exposure time at different positions in the sky, in the
amount of extended cloud emission, and in the source confusion level.  Based on the cumulative
fraction of all sources detected in both epochs of the c2d  observations, the overall 90 
$\%$ completeness levels of the c2d survey have been estimated  to be 0.07, 0.12, 0.5, 0.4, 
and 1.0 mJy  for 3.6, 4.5, 5.8, 8.0, and 24 $\mu$m. See Evans et al. (2005) for a 
detailed discussion of the uncertainties and sensitivity limits of the c2d survey. 

The IRAC observations are sensitive enough to allow robust detections of stellar photospheres in all four IRAC bands for our 
entire sample of wTTs in Ophiuchus and Lupus (distance $\sim$125 pc and $\sim$150--200 pc, respectively) and in $\sim$85$\%$ 
of the objects in Perseus (distance $\sim$320 pc). MIPS 24 $\mu$m observations are not deep enough to reach the stellar
photosphere of some low mass objects (especially in Perseus, which is the most distant cloud we consider), but, in general, 
they are deep enough to detect optically thick disks in our entire sample. The different completeness levels
of our disk census due to sensitivity considerations are discussed in more detail in Section~\ref{completeness}.

\subsection{Sample Selection}\label{sample}

The c2d maps contain several hundred young stars identified by their X-ray and $H\alpha$ 
brightness that have been spectroscopically classified as wTTs stars. Our sample was selected from the 
literature and is distributed as follows: 69 objects in Ophiuchus (Bouvier $\&$ Appenzeller 
1992; Martin et al. 1998), 33 in Lupus (Hughes et al. 1994; Krautter et al. 1997), 
and 130 objects in Perseus (Luhman et al. 2003). The Lupus and Ophiuchus objects are distributed across 
the cloud maps, while the targets in Perseus are located exclusively in the IC 348 cluster. Of the 33 
Lupus objects, 27 are located in the c2d maps of Lupus III and 6 lie within the c2d maps of Lupus I. 

\emph{Spitzer} SEDs of the wTTs in IC 348 have already been presented by Lada et al. (2006, L06 hereafter).
We include these objects in our sample because they  increase the statistical significance of 
our results and allow us to compare a clustered population to the more distributed population
of stars in Lupus and Ophiuchus. Also, we adopt a different disk identification
criterion than L06, which results in a lower disk fraction than that obtained
by the Lada group (see Section~\ref{L06_comp}). 
Furthermore, L06 adopt a single age (2-3 Myrs) for the stars in the IC 348 cluster and  do not attempt 
to study the disk fraction as function of age, which is one of the main goals of 
our paper.

All the objects in our sample, which are listed in Table 1, have known spectral types and small 
$H\alpha$ EWs. The spectral types are necessary to estimate stellar ages from the position of the 
targets in the H-R diagram and the contribution of the stellar photospheres to the observed SEDs, 
 while the $H\alpha$ EWs are required to establish wTTs status. 
The nominal division between cTTs and wTTs is $H\alpha$ EW = 10 $\AA$; however, 
since the $H\alpha$ EW due to chromospheric activity alone can reach this value for late M stars 
(e.g., Martin 1997), we have included in our study 18 M2-M7 stars ( $<$ 8 $\%$ of our 
sample) with $H\alpha$ EW up to 15 $\AA$.  Also, $\sim$8$\%$ of the stars in our
sample show  $H\alpha$ in  absorption rather than in emission. 

We note that even though there is a strong correlation between H$\alpha$ emission and other 
accretion signatures such as optical veiling, a single-epoch low-resolution  measurement 
of H$\alpha$ equivalent width is not enough to rule out active accretion for at least two reasons. 
First, even when a narrow range of spectral types is considered, the distribution of H$\alpha$ 
EWs of  T Tauri stars does not show a clear gap between accreting and non-accreting objects 
(e.g., the $H\alpha$ EWs of weakly accreting PMS stars overlap with those of chromospherically active 
non-accreting stars). Second, accretion itself is a highly variable process, and some objects
constantly move across the wTTs--cTTs H$\alpha$ EW boundary. Therefore, we consider our sample 
to be composed of mostly non-accreting objects but do not rule out the presence of some actively 
accreting interlopers.

\subsection{Complementary data}

In order to construct more complete SEDs of our wTTs, we have collected the 2MASS photometry 
for our entire sample and the V, $R_{C}$, $I_{C}$-band photometry reported by 
Hughes et al. (1994) and  Wichmann et al. (1997) for 24 of our 33 Lupus objects. Also we have obtained our 
own V$R_{C}$$I_{C}$ optical observations for 52 of our 69 Ophiuchus targets and R$_{C}$ and I$_{C}$-band 
observations for 115 of the 130 objects in IC 348 with the 0.8 m telescope at McDonald Observatory. 
The Ophiuchus targets were observed in 7 different 46$\farcm$2 x 46$\farcm$2 fields of view during the 
photometric  nights of June $20^{th}$-$21^{st}$ 2005 with exposures times of 30, 50, and 100 seconds 
for the V, $R_{C}$, and $I_{C}$-band respectively. The objects in IC 348 were observed in a single field of view 
with 200 and 150 second exposures in the R$_{C}$ and $I_{C}$-band, respectively.  In addition to 
the program stars, on each  night, 3 fields of Landolt standards ($\sim$5 standards 
per field) were  observed at different airmasses. The seeing ranged from 1.5$''$ to 2$''$ when the observations were made. 
We reduced the data and perform aperture photometry using the standard IRAF packages CCDRED and DAOPHOT. 
We used a 5.4$''$ (4-pixel) aperture and a sky annulus with inner and outer radii of 16.2$''$ and 22.95$''$ 
respectively. 

The rms scatter of the photometric solutions applied to the programs stars was  $<$ 0.02 mag in all three 
filters. We adopt a conservative \emph{minimum} photometric error of 0.03 mag. We report the magnitudes 
and the uncertainties for all the objects with estimated photometric error less than 
$\sim$0.2  mag. The \emph{Spitzer} photometry for our entire sample is listed in Table 1. 
The non-\emph{Spitzer} data: optical and 2MASS photometry, along with the spectral types 
and H$_{\alpha}$ equivalent widths from the literature, is listed in Table 2.

\section{Results}\label{observations_02}
\subsection{Disk Identification}\label{DF}

In order to identify the stars with disks, we compare the extinction-corrected \emph{Spitzer} 
colors of our targets, to those predicted by NextGen Models (Hauschildt et al. 1999),
convolved with the \emph{Spitzer} bandpasses, for the photospheres of stars of the corresponding spectral types. 
The broader the wavelength baseline of the color used, the larger is the expected excess of the stars with disks; 
therefore, the available color that provides the most clear disk identification is [3.6]--[24]. However, since both 
\emph{Spitzer's} sensitivity and the photospheric fluxes decrease with increasing wavelength, not all sources are 
detected at wavelengths longer than 5.8 $\mu$m. 

In Figure 1a, we plot  [3.6] vs EX([3.6]--[24]), for the 98 stars with 3.6 
and 24 $\mu$m fluxes available, where EX([3.6]--[24]) is ([3.6]--[24])$_{OBSo}$-([3.6]--[24])$_{Model}$,
([3.6]--[24])$_{OBSo}$ are the observed colors corrected for extinction, and ([3.6]--[24])$_{Model}$
are the photospheric colors predicted by the NextGen Models. We estimate the extinction, $A_V$, 
using $A_{V}= 4.76 \times E(R-I) = 4.76\times((R_{C}-I_{C})_{OBS}-(R_{C}-I_{C})_{O})$. 
The intrinsic stellar colors, (R$_{C}$-I$_{C}$)o, come from Kenyon \& Hartmann (1995). For objects without 
R$_{C}$ and I$_{C}$ fluxes available, we use $A_V$=5.88$\times$E(J-K$_{S}$). We note that this will result 
in an overestimated extinction for objects with significant K-band excess. The extinction at 8 and 
24 $\mu$m are calculated according to Table 3. 

We consider objects with  [3.6]--[24] $<$ 0.7 to be those without excess whose emission arises solely from the
stellar photosphere. The mean EX([3.6]--[24]) value for this group is not zero but 0.07 mag  with a 1-$\sigma$ 
dispersion of 0.17 mag. The 0.07 mag offset of the observed stellar photospheres with respect to the
models is probably due to a combination of the systematic errors in the absolute flux calibrations, the stellar models, 
and the extinction corrections. In  Figure 1a, we treat this offset by subtracting 0.07 mag from all the EX([3.6]--[24]) values. 
We find that  40 objects have a [3.6]--[24] excesses larger than 5-$\sigma$. These are very robust disk identifications. 
One object, RXJ1622.6-2345, has [3.6]--[24] excess just over 3-$\sigma$. We consider this object to be a good disk 
candidate, but warn the reader of its lower significance. Since the c2d MIPS maps cover a larger area than the IRAC maps, 
24 $\mu$m is the only \emph{Spitzer} flux available for 5 stars in our sample. In these cases, we use [K]--[24] colors for disk
identification. None of the 5 objects for which MIPS 24 $\mu$m is the only available \emph{Spitzer} flux show a significant
K--24 $\mu$m excess. Of the 127 stars without measured [3.6]--[24] or [K]--[24] colors, 112 have [3.6]--[8.0] colors available. 
Figure 1b is analogous to Figure 1a, but here we plot  [3.6] vs EX([3.6]--[8.0]) for all the stars with measured [3.6]--[8.0] colors, 
including the ones from Figure 1a, which are shown as open diamonds. 
Following Cieza $\&$ Baliber (2006), objects with [3.6]--[8.0] $<$ 0.7 are considered stellar photospheres.
In this case, the mean color offset with respect to the models is 0.05 mag. The standard deviation of the stellar 
photospheres is 0.16 mag, but the error clearly increases with decreasing brightness. We find that only 6 objects 
not detected at 24 $\mu$m show a clear evidence ($>$ 5-$\sigma$) for 8 $\mu$m excess.  Two objects, IC348-76 and IC348-67, 
show excesses between 3 and 5-$\sigma$. Since the SEDs of these two objects show a hint of IR-excess at 4.5 and 5.8 $\mu$m,
we consider these two objects to be good disk candidates. We find that none of the 15 objects without [3.6]--[24], 
[K$_{S}$]--[24], or [3.6]--[8.0] colors available shows a significant [3.6]--[5.8] excess. 

The SEDs of the wTTs disks in Lupus, Ophiuchus, and IC 348 are shown in Figures 2, 3, and 4, respectively.
The open squares represent the observed optical, 2MASS, IRAC and MIPS-24 $\mu$m fluxes, while the dots correspond 
to the extinction corrected values. The A$_{V}$'s are estimated as described above for Figure 1, while the extinctions 
at other wavelengths are calculated according to Table 3.
NextGen model photospheres, corresponding to published spectral types and normalized to the extinction-corrected J-band, 
are shown for comparison.

\subsubsection{Disk Census Completeness}\label{completeness}

Following the procedure described above, we identify a total of 46 wTTs disks and 3 disk candidates.
For definiteness, we consider these 3 disk candidates to be real in the rest of the paper.
However, we note that the completeness of our disk census is lower for our IC 348 sample than it is for the  sample of Lupus and 
Ophiuchus wTTs. 

The IRAC observations are sensitive enough to allow robust detections of stellar photospheres in all four IRAC bands for 
wTTs in the sample from Lupus and  Ophiuchus (102 objects in total). Also, the 5 objects in Lupus and Ophiuchus that fall outside the 
c2d IRAC maps all have 24 $\mu$m  fluxes consistent with stellar photospheres, which rules out the presence of any significant 
IRAC excess in these objects. Therefore, the census of IRAC excesses in our sample of Lupus and Ophiuchus wTTs is likely to 
be complete (i.e. only 17 out of the 102 wTTs in Lupus and Ophiuchus have significant IRAC excesses). 
At 24 $\mu$m, we detect \footnote{by  detect 
we mean that the flux of the object has been measured regardless of whether or not the star has an excess. All the objects have been 
observed at 24 $\mu$m;therefore, a non detection implies that the flux of the object is below our 24 $\mu$m sensitivity limit.} 
80$\%$ (82/102) of the Lupus and Ophiuchus objects.  Thus, it is possible that some of the 20 wTTs from the Lupus and Ophiuchus sample 
that are not detected  at 24 $\mu$m have SEDs that start to diverge from their photospheres longward of 8.0 $\mu$m but remain bellow our
sensitivity at 24 $\mu$m. We find that all the objects with IRAC excess are detected at 24 $\mu$m, and 
that only 2 of the 82 objects detected at 24 $\mu$m have excesses that only become evident at this wavelength. 

In our IC 348 sample, 11$\%$ (14/130) of the objects are not detected at 8.0 $\mu$m and 82 $\%$ (107/130)
are not detected at 24 $\mu$m. The  longer integration times of the GTO IRAC observations 
of IC 348 with respect to the c2d observation of Lupus and  Ophiuchus (1548 vs. 46.8 secs)  more 
than compensates the effect of the distance (320 vs. 125--200 pc) on the expected sensitivities. 
However, the high background of the IC 348 cluster becomes the limiting factor for the detection 
of faint sources at 8.0 $\mu$m. At 24 $\mu$m, the depth of the GTO observations only partially compensates 
for the greater distance of IC 348, and the effect of the background becomes even larger than it is at  
8.0 $\mu$m. The combination of these two factors explains the very low detection rate of IC 348 members at 24 $\mu$m. 
None of the 14 objects without a 8.0 or 24 $\mu$m \emph{detection} shows a significant 5.8 $\mu$m excess
that would indicate the presence of a circumstellar disks. However, some of these 14 wTTs 
could have SEDs that start to diverge from their photospheres longward of 5.8 $\mu$m but remain bellow
our sensitivity at 8.0 and 24 $\mu$m. Finally, we find that 91$\%$ (21/23) of the objects detected at 24 $\mu$m 
also show significant IRAC excesses, which is consistent with the results for Lupus and Ophiuchus. Obviously, 
in IC 348, we are likely to have missed most of the stars for which the onset of the IR-excess occurs longward 
of 8.0 $\mu$m.   

\subsubsection{Disk fraction statistics}\label{DF_stats}

In section \ref{DF} we tried to present a disk census as complete as possible given the available 
data. However, since the sensitivity of our disk survey is not uniform across all wavelengths and 
varies from region to region due to distance and background level effects, we derive disk fraction 
statistics from the number of objects with excess at IRAC wavelengths (IRAC disk fraction, hereafter) 
in a magnitude limited subsamples that is not likely to be affected by sensitivity variations.  

As discussed in the previous section, we consider our census of IRAC excesses in our sample of
Lupus and Ophiuchus wTTs to be complete. Excluding the disks that are only detected longward of 
10 $\mu$m (ROXs 36, and RX J1622.6-2345), we derive the following IRAC 
disk fractions:  27$\%$$\pm7\%$ (9/33) for Lupus, 13$\%$$\pm4\%$ (9/69) for Ophiuchus. 
For IC 348, we restrict our sample to the 96 objects with 3.6 $\mu$m fluxes greater 
than 3.2 mJy, a level above which we also detect at 8.0 $\mu$m every object in our sample. 
We find that 22 of these 96 objects show significant IRAC excess, which leads to
a disk fraction of 23$\pm$4$\%$. Combining the objects in Lupus, Ophuichus, and IC 348,  we derive 
an overall IRAC disk fraction of 20$\pm$3$\%$ (40/198). The IRAC disk fractions of wTTs in Lupus 
and Ophiuchus bracket that of the IC 348 cluster. This suggest that the disk fraction in IC 348 wTTs
targets is not strongly affected by the cluster environment.

At 24 $\mu$m, the sensitivity of our survey is significantly less uniform than at IRAC wavelengths. 
We find that even some of our brightest objects lack 24 $\mu$m detections due to the strong 
background emission surrounding them. This prevent us from deriving robust MIPS disk fractions 
statistics even from a magnitude limited subsample. 

\subsection{Disk Properties}\label{disk_properties}

\subsubsection{Selected Color-Color Diagrams}

In this section,  we place our sample of wTTs in \emph{Spitzer} color-color
diagrams as our first attempt to explore the diversity of wTTs disks. For comparison, we also include
in these diagrams two samples of 83 and 66 cTTs from Taurus and IC 348, respectively. 
We used the H$\alpha$ EW from Luhman et al. (2003) to select the sample of cTTs in IC 348.
For these objects we use our own photometry. The \emph{Spitzer} fluxes of the Taurus cTTs come 
from Hartmann et al. (2005). In Figures 5 and 6, we show an $A_{V}$=10 extinction 
vector based on Table 3; however, we note that for most of the stars in our sample the extinction 
is \emph{significantly} smaller. The mean A${_V}$ we derive for our Lupus, 
Ophiuchus, and IC 348 wTTs samples are 0.64, 2.7, and 1.7 mag respectively. Also, 93$\%$ of 
our sample has A$_V$ $<$ 5 mag, and the remaining 7 $\%$  has 10 mag $<$ $A_{V}$ $<$  5 mag. 

Figure 5a shows the [3.6]--[24] vs. [3.6]--[8.0] colors of our sample of wTTs stars.   
Based on the colors shown, we identify three different groups. The first group consists 
of objects with [3.6]--[24] $<$ 0.7 and [3.6]--[8.0] $<$ 0.4, which are consistent with bare stellar photospheres. 
The second group consists of stars with [3.6]--[24] $>$ 0.7 and [3.6]--[8.0] $<$ 0.4 (i.e. their SEDs start to diverge
from their stellar photospheres longward of 8.0 $\mu$m). The 4 stars in this group are RXJ1622.6-2345, ROXs 36, 
IC348-56, and IC348-124. In Section~\ref{FDL}, we find that the objects in this second group  have 
the lowest fractional disk luminosities of the sample. In fact, all four objects have $L_{DISK}$/$L_{*}$ $<$ 10$^{-3}$, which suggest 
the presence of optically thin disks. 
The last group of objects has [3.6]--[24] $>$ 0.7 and [3.6]--[24] $>$ 0.4 (i.e. they show excess at both IRAC and MIPS 
wavelengths). Most of these objects are likely to be optically thick disks.  Figure 5b shows our sample of wTTs combined with the 
sample of cTTs. We note that the cTTs populate exclusively the region of the diagram we associate  with optically thick disks
and that both wTTs and cTTs  populations are very well mixed in this region of the diagram.  

In Figure 6a, we show the [3.6]--[4.5] vs. [5.8]--[8.0] colors of wTTs. In general,  the populations of stars with 
and without an excess are clearly separated. Stars in the upper right corner of the diagram have both 4.5 
and 8.0 $\mu$m excesses. Stars in the lower right corner of the diagram are stars with 8.0 $\mu$m excess but 
no 4.5 $\mu$m excess. These objects are usually interpreted as ``transition disks'' with inner holes
(e.g. Calvet et al. 2002; D'Alessio et al. 2005). In fact, Sicilia-Aguilar et al. (2006) define transition disks as 
objects  with 8.0 $\mu$m  excess but no 4.5 $\mu$m excess. They  find that only $\sim$10$\%$ of the T Tauri disks in 
the 4-Myrs old cluster Tr 37 fall into this category.  In contrast, we find that $\sim$30 $\%$ of the wTTs are in 
fact ``transition disks'' according the working definition stated above. This is in good agreement with the idea that wTTs 
disks represent a more evolved evolutionary state than cTTs disks. In Figure 6b, we combine wTTs and cTTs in the same 
color-color diagram. 

\subsubsection{$EW(H_{\alpha})$ dependence?}\label{DF_vs_Halpha}

Since the distinction between cTTs and wTTs is based on the  H${_\alpha}$ EW, and cTTs show a
much higher disk fraction than wTTs, we investigate the dependence of the IRAC disk fraction on 
$H\alpha$ EW within our sample of wTTs. We restrict our analysis to the 198 objects in the magnitude 
limited subsample discussed in Section ~\ref{DF_stats}. Figure 7a shows a histogram of the $H\alpha$ EW 
of stars with and without a disk, while Figure 7b shows the disk fraction vs. $H\alpha$ EW. We find that (1) 
the disk fraction is highly correlated with $H\alpha$, and (2) the disk fraction is a smooth function of 
$H\alpha$ EW, increasing from $\sim$5$\%$ for the stars with $H\alpha$ observed in absorption to $>$50 $\%$ 
for the stars with 15 $\AA$ $>$ $H\alpha$ EW $>$ 10 $\AA$.

Since a strong correlation between the presence of a disk and H$\alpha$ emission of chromospheric origin  
would be difficult to explain, the correlation observed in Figure 7 is likely to be driven 
by the presence of weakly accreting stars in our sample of wTTs.  Muli-epoch high-resolution spectroscopy of
the sample of wTTs with disks would be highly desirable to establish, from their H$\alpha$ velocity profiles, 
which objects are in fact actively accreting (White $\&$ Basri 2003). 

\subsubsection{Disk fraction dependence on spectral type?}

Lada et al. (2006) find that the ``optically thick'' disk fraction in the entire populations of PMS stars (cTTs/wTTs) 
in IC 348 seems to be a function of spectral type. They argue that the disk fraction peaks around K6--M2 stars, which 
at the age of the cluster corresponds to stars with  masses similar to that of the Sun, and conclude that circumstellar 
disks might last longer around solar type stars than around  both less and more massive stars. Similarly, Carpenter et al. 
(2006) studied a sample of over 200 PMS star in the 5 Myrs old Upper Scorpius OB association with masses between $\sim$0.1 
and 20 M$_{\odot}$ and found that stars with K-M spectral types have a significantly larger disk fraction than stars with G-B 
spectral types. 

In order to investigate the possibility of a disk fraction dependence on spectral type for wTTs, we restrict our analysis 
to the 198 objects in the magnitude limited subsample discussed in Section ~\ref{DF_stats} and divide our sample into the 
same 3 spectral type bins studied by Lada et al. (2006). We find the following IRAC disk fractions for the 3 bins: 28$\%$$\pm7\%$ 
(12/42) for stars with spectral types  K6 and earlier, 16$\%$$\pm5\%$ (10/62) for K7--M2 stars, and  19$\%$$\pm4\%$ (18/94) 
for M2--M6 stars. Our result is inconsistent with a disk fraction of wTTs peaking around  K6-M2 stars
and suggests that the disk fraction dependence on spectral type found by Lada et al. 2006, if real,  is likely to be 
driven by the cTTs population, which would imply that the duration of the \emph{accretion phase} is a function 
of spectral type. Given the very different spectral type distribution of our sample (there are only 7 objects with
types later than K0 in our sample), our result does not contradict those presented by Carpenter et al. (2006).  

\subsubsection{Fractional Disk Luminosity}\label{FDL}

The ratio of the disk luminosity to the stellar luminosity, $L_{DISK}/L_{*}$, is a measurement of the fraction of 
the star's radiation that is intercepted and re-emitted by the disk plus any accretion luminosity. 
This quantity is intimately related to the evolutionary status of a circumstellar disk. 
On the one hand, the primordial, gas rich, disks around cTTs have typical
$L_{DISK}/L_{*}$ values $>10-20$$\%$. On the other hand, $L_{DISK}/L_{*}$ values for optically 
thin, gas poor,  debris disks around MS stars range from $10^{-3}$ to $10^{-6}$ (Beichman et al. 2005). 
In order to characterize $L_{DISK}/L_{*}$ for wTTs disks, we 
estimate this quantity for all the wTTs disks in our sample except for the few disks in IC 348 for which 
24 $\mu$m fluxes are not available. These objects are below our 24 $\mu$m sensitivity limits, and the disk 
luminosities obtained from the IRAC excesses alone would be highly uncertain.

We estimate the disk luminosity according to the following procedure. We first calculate the IR-excess 
at IRAC wavelengths and  at 24 $\mu$m by subtracting from the observed fluxes the expected photospheric 
contributions predicted by the stellar models shown in Figures 2, 3, and 4. Then, the IR-excess at 24 
$\mu$m is extrapolated to longer wavelengths assuming the emission of a black body peaking at 24 $\mu$m
(i.e., T = 121 K) and diluted by an emissivity proportional to $\lambda$$^{-1}$.  Also, the IR emission of 
the shortest \emph{significant} IRAC excess is extrapolated to shorter wavelengths assuming black body 
emission. The temperature of the adopted black bodies ranges from 1400 to 500 K depending on the IRAC band 
from which the IR-excess is extrapolated. Finally, we calculate the total disk luminosities by integrating the 
observed and extrapolated IR-excesses over frequency. Similarly, the stellar luminosities are
calculated by integrating the fluxes of the stellar models shown in Figures 2, 3, and 4 over 
frequency. 
The distribution of $L_{DISK}/L_{*}$ for our sample of wTTs is shown in Figure 8.  
The  $L_{DISK}/L_{*}$ values calculated using the same procedure for a sample of cTTs (From Cieza et al. 2005) and 
debris disks (from Chen et al. 2005) are shown for comparison.
The $L_{DISK}/L_{*}$ values we derive for our sample should be considered lower limits because our assumption that 
the flux density peaks at 24 $\mu$m may underestimate the flux contribution of cool material in the outer disk.
However, since none of our disks are detected at 70 $\mu$m, we can constrain the flux contribution of the outer 
disks that may be missing in our disk luminosity estimates. For each object, we approximate the outer disk as a 
diluted black body peaking at 70 $\mu$m  (i.e., T = 41 K) with a flux density equal to 45 mJy, the estimated 
3-$\sigma$ limits of the c2d observations at 70 $\mu$m (Evans et al. 2005). For objects in Lupus and Ophiuchus, this approach 
constrains the luminosities of the outer disks to the $10^{-5}$--$10^{-4}$ $L_{*}$ range. For objects in IC 348, 
this range is $10^{-3}$--$10^{-4}$ $L_{*}$. Thus, we conclude that, for objects with warm disks for which 
we derive $L_{DISK}/L_{*}$ $>$ $10^{-3}$, the flux contribution of the outer disks is a small fraction of the total disk 
luminosity. For the objects we derive $L_{DISK}/L_{*}$ $<$ $10^{-3}$, the contribution of an unseen outer 
disk could, in principle, dominate the total disk luminosity. However, the disk luminosities of such cold disks 
would still remain in the debris disk range. 

Figure 8 shows that only one object, Sz 96, has a $L_{DISK}/L_{*}$ value characteristic of cTTs. This 
object is a M2 border line cTTs/wTTs with a $H\alpha$ EW of 11 $\AA$. As shown in Figure 2, the SED of Sz 96 
shows strong IR-excess at 2MASS wavelengths and is indistinguishable from that of a cTTs. This object is likely to be
one of the actively accreting interlopers discussed in Sections \ref{sample} and \ref{DF_vs_Halpha}.
Similarly, only 4 objects have $10^{-3} > L_{DISK}/L_{*}  >  10^{-6}$, the range characteristic of optically thin debris disks. 
However, we note that the low luminosity and low optical depth of a disk do not necessarily warrant debris disk status, which 
requires the presence of second generation of dust produced by the collision of planetesimals in a gas poor environment. 
The lowest luminosity, and presumably optically thin, disks in our sample are RXJ1622.6-2345, ROXs 36,  IC348-56, 
and IC348-124. We model the SEDs of these disks in Section~\ref{DISK_MODELS}. If these objects are in fact debris disks, at an 
age of $\sim$1--3 Myrs, they could be some of the youngest debris disks observed to date. However, the confirmation of debris 
disk status would require information on the grain size distribution and gas content of their disks. The main conclusion that 
can be drawn from Figure 8 is that the bulk of the wTTs disks have $L_{DISK}/L_{*}$ values that bridge the 
gap between the cTTs 
and the debris disk range. This supports  the idea that wTTs disks represent an intermediate evolutionary stage linking 
primordial disks around cTTs and debris disks 
around MS stars. 

\section{Discussion}
\subsection{Comparison to recent Spitzer results}
\subsubsection{Comparison to Padgett et al. 2006}\label{PO6A}

      The IRAC disk fractions we found in Section~\ref{DF} are $\sim$3--6 times larger than those 
recently found for a sample of 83 relatively isolated wTTs (Padgett et al. 2006, P06 hereafter). 
P06 find that 3 of their objects show IR-excess both at IRAC and MIPS wavelengths, while 2 objects
show IRAC fluxes consistent with photospheric emission and small excesses at 24 $\mu$m. Most of these 
stars in the P06 sample are members of the extended population of Li rich wTTs discovered by the 
ROSAT X-ray satellite around nearby molecular clouds and are, for the most part, located outside  
the high extinction regions mapped by the c2d project, but within $\sim$6 degrees of the centers of the 
c2d cloud maps.The P06 observations were sensitive to the stellar photospheres of the entire sample
at both IRAC wavelengths and 24 $\mu$m. Given the size of the samples involved, the discrepancy 
in the IRAC disk fraction of the off-cloud wTTs studied by P06, 4$\pm2 \%$, and that of 
the on-cloud wTTs studied in this paper, 20$\pm3\%$, is significant at the 4-$\sigma$ level. 
  
       Follow up spectroscopic studies of ROSAT targets (e.g. Covino et al. 1997; Martin et al. 1999; Wichmann et al. 2000) 
used the strength of the Li I 6707 absorption line to discriminate bona-fide PMS from young active MS stars (possibly $\sim$100 
 Myrs old) sharing the same X-ray properties. P06 selected their sample from ROSAT wTTs with  Li I 6707 absorption lines 
stronger than that of a Pleiades star of the same spectral type. These objects were suspected to be 1--10 Myr-old because the 
age distribution of the ROSAT wTTs, derived from their position in the H-R diagram, is not significantly different from that 
derived for the on-cloud TT Tauri stars (Alcala et al. 1997; see also Section~\ref{DF_vs_age} of this paper).
This age estimate assumes that the ROSAT sources are located at the same distances as their adjacent molecular clouds.       
       However, the distance and pre-main sequence status of the extended population of ROSAT sources has been called 
into question. Feigelson (1996) and Briceno et al. (1997) argue that the distributed population of ROSAT 
sources consist of mostly foreground post-T Tauri and active young ($\sim$100 Myrs old) MS stars which are not 
necessarily associated with regions of current star formation (i.e. they are old enough to have moved far away from their 
birth sites) and that the presence of Li I 6707 does 
not warrant PMS status since it is highly dependent on stellar mass, convection, and angular momentum history. 
       The low disk fraction of the distributed population of ROSAT sources studied by P06 with respect to that of the 
sample of on-cloud wTTs studied here supports the idea that the ROSAT sources found \emph{around} molecular clouds 
represent a significantly older population of stars than that of the $<$ 1-10 Myrs old cTTs and wTTs found \emph{within} 
molecular clouds. 
       The age estimate of the extended population of Li rich ROSAT sources depends on whether or not the 
P06/ROSAT targets are at the same distances as their adjacent molecular clouds. The luminosities of stars closer to us than 
the molecular clouds themselves could be considerably overestimated,  
translating  into underestimated ages when the stars are placed in the H-R diagram. 
If the P06 sources are in fact associated with the molecular cloud and the sample has been biased toward foreground stars, 
the wTTS in P06 sample could be somewhat older than their  nominal age of 1--10 Myr. It is reasonable to assume that such a bias
toward foreground objects exists because P06 favored bright objects when selecting their sample in order to increase the number of wTTs 
that could be observed at a given sensitivity. If the P06 sources are not associated with their adjacent molecular clouds, 
then the only age constraints are Li I 6707 EW and X-ray brightness, and ages of the order of 100 Myrs could not be 
ruled out.  

\subsubsection{Comparison to Lada et al. 2006}\label{L06_comp}

In a recent study of $\sim$300 confirmed members of IC 348 (wTTs and cTTs) Lada et al. (2006, L06 hereafter) 
find an IRAC disk fraction of $\sim$35$\%$ among IC 348 members classified as wTTs. The sample of 
wTTs studied by L06 is virtually the same sample of IC 348 wTTs included in this paper, 
for which we derive an IRAC disk fraction of 23$\%\pm4\%$. In what follows, we attempt to account 
for the difference in these results. L06 use a disk identification criterion based on 
the slope, $\alpha$,  of a power law fit to the four IRAC bands.  From the comparison to disk models, 
they identify objects with $\alpha$ $>$ --1.8 as optically thick disks.  Based on the predicted slope of 
an M0 star and the typical uncertainty in the power law fit, they identify objects with $\alpha$ $<$ --2.56 
as stellar photospheres and objects with intermediate slopes, --2.56\,$>$\,$\alpha$\,$>$\,--1.8. 
as ``anemic disks''. Cieza $\&$  Baliber (2006) suggest that many of the objects identified as ``anemic disks''
by the L06 are more consistent with stellar photosphere of late M stars  than with circumstellar disks. 
Based on a large sample ($>$ 400) of cTTs and wTTs from the literature with photometric uncertainties smaller 
than 0.1 mag,  Cieza $\&$ Baliber (2006) find that cTTs and wTTs disks occupy a  well defined locus in the [3.6]--[8.0] 
vs [3.6]--[5.8] diagram with  [3.6]--[8.0] $>$ 0.7, while bare stellar photospheres have [3.6]--[8.0] $<$ 0.5. 
In fact, only $\sim$1 $\%$ of the their sample have 0.7 $>$ [3.6]--[8.0] $>$ 0.5. As seen in  Figure
6, faint objects in IC 348, most of which are M4--M7 stars, have very uncertain [3.6]--[8.0] colors. 

We suggest the possibility that the combined effect of slightly redder stellar photospheres and larger photometric 
uncertainties in late M stars with respect to brighter stars with earlier spectral types resulted in a large fraction 
of M4--M7 stellar photospheres being classified as ``anemic'' disks by L06. L06 find that the ratio of 
``anemic'' disks to optically thick disks increases from $\sim$1/5 to $\sim$1/1 from K6--M2 to M2--M6 stars,
which is consistent with the idea that some of the anemic disks around late type stars
may be of questionable significance. 

We find that most of the wTTs that do not satisfy our disk identification criterion discussed in Section~\ref{DF}
but are classified as ``anemic'' disks by L06 are objects with spectral  types M5 or later, which tend to have very 
uncertain IRAC colors due to the strong IR background of the IC 348 cluster. These low-mass objects account for 
the difference in the disk fraction derived by us and L06 for the wTTs in IC 348. In Table 4 we list the 18 objects 
classified as anemic disks by L06 that do not satisfy our disk identification criterion discussed in Section~\ref{DF}. 
We find that, for objects in Table 4, the [3.6]--[8.0] colors obtained by L06 are $\sim$0.15 mag redder in the mean 
than our colors. However, this is a selection effect due to the fact that 
we are only considering the objects that have [3.6]--[8.0] colors from L06 that are red enough to be classified as 
anemic disks. In fact, when the entire sample of IC 348 wTTs is considered, our photometry gives [3.6]--[8.0] colors
that are $\sim$0.05 mag \emph{redder} than those obtained by L06. We note that even using the colors from L06, 
only one of the objects satisfies [3.6]--[8.0] $>$ 0.7,  the disk identification criterion proposed by Cieza \&  
Baliber (2006). Also, for 12 of the 18 objects listed in Table 4, L06 derive $\alpha$'s that are consistent with 
the expected stellar photosphere within the errors (i.e. ($\alpha$+2.66)/$\sigma$$_{\alpha}$ $<$ 3, where -2.66
is the slope of an M0 photosphere adopted by L06). Finally, two of the objects for which 
($\alpha$+2.66)/$\sigma$$_{\alpha}$ $>$ 3 have no 8.0 $\mu$m fluxes measured by us or by L06, which renders the disk 
identification less reliable. We conclude that an accurate identification of weak IRAC excesses requires careful 
consideration of the photometric uncertainties involved, especially in regions with strong extended emission such
as the IC 348 cluster.

\subsection{The diversity and evolutionary status of wTTs disks}

For late type stars, the IRAC and MIPS fluxes originate from the inner
$\sim$20 AU of the disk. For this reason, observations at these wavelengths are
especially suited to study the evolution of the planet-forming region of the disk.   
According to models of the evolution of circumstellar disks, dust growth and 
settling take place very efficiently in the circumstellar disks soon after 
their formation (Weidenschilling 1997, Dullemond \& Dominik 2005). 
Numerical models also predict that the growth of dust will reduce the opacity of 
the disk to the stellar radiation causing the disk to become flatter and the
excesses in the mid infrared to decrease with time proceeding from the
shortest to the longest wavelengths. There is mounting evidence of
such phenomena in the IRAC colors of T Tauri stars and lower mass
objects (D'Alessio et al. 2001, Furlan et al. 2005, Sicilia-Aguilar et
al. 2005, Muzerolle et al. 2006).
                                                                                                                
In this section, we try to explore observational consequences that
could be related to dust grain growth
and/or settling toward the mid-plane for the sample of wTTs with
disks. To this end, we estimate the longest wavelength \emph{without}
significant infrared excess from the dereddened SEDs, $\lambda_{\rm
turn-off}$ (in $\mu$m), and the slope of the SED longward of
$\lambda_{\rm turn-off}$, $\alpha_{\rm excess}$ for all the wTTs stars
with disks in the sample. We compare these values for our sample of
wTTs to those we obtain for a sample of cTTs in Chamaeleon from Cieza et al. 
(2005) and a sample of debris disks from Chen et al. (2005).
Figure 9 shows the ranges of mid-IR slopes $\alpha_{\rm
excess}$ versus the wavelength of disk emission turn off
$\lambda_{\rm turn-off}$. Systems with $\lambda_{\rm turn-off}$ $\le$
2.0  $\mu$m show some excess in the near-IR JHK bands and are probably
actively accreting (see e.g. Hartmann et al. 2005 and Muzerolle et al. 2003). 
Objects with $\lambda_{\rm turn-off}$ in the IRAC range are likely to be purely
irradiated disks with increasing degree of settling and/or larger
inner holes with increasing $\lambda_{\rm turn-off}$. Finally, those stars
with $\lambda_{\rm turn-off} \ge$ 8.0 $\mu$m are likely to have cleared 
inner disks with radii of several AU (e.g. Calvet et al. 2002) 
or alternatively have dust in the inner disk which has grown to
sizes large enough not to produce any detectable excess above the
photospheric level.
                                                                                                                
A simple disk evolution scenario based solely on dust grain growth and
settling predicts smaller $\alpha_{\rm excess}$ slopes for longer
$\lambda_{\rm turn-off}$ or larger cleared inner regions (Dullemond \&
Dominik 2005, D'Alessio et al. 2005). However, Figure 9
suggests that the {\it range of possible excess slopes increases with
the process of inner disk clearing}. The actively-accreting cTTs stars
all cluster around $\alpha_{\rm excess}$ values of $-1$. For these objects,
the onset of the IR excess occurs near 2.2 $\mu$m.
We find that  wTTs show a richer  distribution of SED morphologies
than cTTs.  Some of the wTTs show SEDs that are indistinguishable 
from those of cTTs, while others show a wide range of $\lambda_{\rm 
turn-off}$  $\alpha_{\rm excess}$ values. Finally, the debris disks, 
with ages 12 and 200 Myrs (Chen et al. 2005) show detectable excess 
only longward of 8.0 $\mu$m, but also present a great variety of spectral 
slopes.
                                                                                                                
In general, the diagram suggests an evolutionary sequence in which most
actively accreting cTTs have similar near to mid-IR SEDs, dominated by
optically thick emission of the inner disk. This similarity is probably 
due to the fact that, in most cases, their inner disks extend to the 
dust sublimation radius (Muzerolle et al. 2003). On the other hand, 
if disks evolve from the inside out, wTTs disks are likely to have a wider 
range of of inner disk radii and temperatures than cTTs. This diversity of inner 
disks, together with possible grain grown and grain settling to the disk mid-plane 
may explain the large ranges of $\alpha_{\rm excess}$ and $\lambda_{\rm turn-off}$ found. 
Finally, the debris disks populate the right part of the plot with large cleared inner disks and 
emission from cool optically thin dust. 

\subsection{Comparison to Optically Thin Disk Models}\label{DISK_MODELS}

As discussed in Section~\ref{disk_properties}, four stars in our sample show SEDs 
and L$_{DISK}$/L$_{*}$ consistent with optically thin disks. These are ROXs 36, 
RXJ1622.6-2345, IC348-56  and IC348-124.In this section, we modeled these four objects 
using the the optically thin disk model developed by Augereau et al. (1999). 

We limited the exploration of the parameter space to the disk parameters that affect most the 
global shape of an SED, namely the minimum grain size $a_{\rm min}$, the peak surface density 
position $r_0$ and the total dust mass $M_{\rm dust}$ (or, equivalently, the surface density at $r_0$).
We adopted a differential grain size distribution proportional to $a^{-3.5}$ between 
$a_{\rm min}$ and $a_{\rm max}$, with $a_{\rm max} = 1300\,\mu$m, a value sufficiently large 
to not affect the SED fitting in the wavelength range we consider. 
Following Augereau et al. (1999), the disk surface density $\Sigma(r)$ is  parametrized by a two power-law 
radial profile $\Sigma(r) = \Sigma(r_0) \sqrt{2} \left(x^{-2\alpha_{\rm in}}+x^{-2\alpha_{\rm out}} \right)^{-1/2}$ 
with $x = r/r_0$, and where $\alpha_{\rm in} = 10$ and $\alpha_{\rm out} = -3$ to 
simulate a disk peaked around $r_0$, with a sharp inner edge, and a density profile decreasing 
smoothly with the distance from the star beyond $r_0$. The optical properties of the grains 
were calculated for astronomical silicates (optical constants from Weingartner $\&$ Draine (2001)), 
and with the Mie theory valid for hard spheres. The grain temperatures were obtained by assuming the
dust particles are in thermal equilibrium with the central star. NextGen model atmosphere spectra 
(Hauschildt et al. 1999) scaled to the observed dereddened K-band magnitude, were used 
to model the stellar photospheres.
                                                                                                              
For each star, we calculated $15000$ SEDs ($0.3\,\mu$m$\leq \lambda \leq 950\,\mu$m),
for $75$, logarithmically-spaced values of $a_{\rm min}$ between $0.05\,\mu$m and
$100\,\mu$m, and for $200$ values of $r_0$, logarithmically-spaced between $0.02$\,AU
and $300$\,AU. For each model, the dust mass was adjusted by a least-squares method,
assuming purely photospheric emission (within the uncertainties, $10\%$) in the four
IRAC bands, and by fitting the measured MIPS $24\,\mu$m flux density. Models with flux densities 
in the MIPS $70\,\mu$m and $160\,\mu$m bands, larger than the  estimated $2\sigma$ upper limits 
of $0.03\,$Jy and $0.5\,$Jy, respectively, were eliminated. 
The results are summarized in Table 5, and the SEDs are displayed in Figure  10. 
Results in Table 5 are listed for two different regimes of minimal grain sizes, 
namely, $a_{\rm min}$ $>$ 10 $\mu$m and $a_{\rm min}$ $<$ 0.5 $\mu$m. 
The first regime accounts for a scenario where all grains have grown
to sizes $>$ 10 $\mu$m and where smaller grains are not replenished by
collisions, while the second regime accounts for a case where grains
smaller than 0.5 $\mu$m remain in the system, either because they are
collisionally replenished, or because pressure forces are inefficient
to expel grains. However, assuming a gas poor environment (i.e. the dust dynamics is 
not controlled by the gas), dissipation processes acting on small grains will limit the 
size of the smallest grains that are likely to populate the disk. For ROX 36, an A2 star, 
the blow-out size is of the order of a few microns. Thus, grains significantly smaller 
than 0.5 $\mu$m are unbound and unlikely to be present in the disk. For the three M-type stars, 
radiation pressure is not high enough to overcome the gravitational force and expel grains. 
In this case, and provided the star mass loss rate is high enough,
the pressure force expected from the stellar wind will set the lower
limit for $a_{\rm min}$ (Augereau $\&$ Beust 2006, Strubbe $\&$ Chiang 2006).

In most cases, neither the position of the peak surface density $r_0$, nor the minimum 
grain size $a_{\rm min}$, can be uniquely determined with so few observational constraints, 
but some models can be eliminated. In particular, the lack of excess in the IRAC bands 
imposes the disk to be significantly dust-depleted within $\sim$1 AU from the star.
The best fits to the SEDs of the three low-mass stars 
(RXJ1622.6-2345, IC348-56  and IC348-124)
are indeed obtained for $r_0$ values around $2$--$5$\,AU, while in the case of ROXs 36, 
the inner hole could be ten times larger. 
The properties of the ROXs 36 disk are in fact those of a typical $\beta$ Pic-like 
disk: a dust mass in the $10^{-4}$--$10^{-2}$\,M$_{\oplus}$ range, a low fractional 
luminosity $L_{\rm IR}/L_*$ of about $10^{-4}$, and typical minimum collision time-scales 
(as calculated using the formula given by Backman et al. 1993) one to three orders 
of magnitudes smaller than the star age. Also, as it is the case for virtually 
all $\beta$ Pic-like disks, the disk around ROXs 36 seems to be collisional 
dominated because collisions occur much faster than Poynting-Robertson (P-R) drag 
(as calculated using the formula given by Augereau $\&$ Beust 2006) for all grains 
larger than the blow-up size.
The best fits to the observed excesses around the three low-mass
wTTs (RXJ1622.6-2345, IC348-56 and IC348-124) are also consistent with optically 
thin disks observed around other young M-type stars such as AU Mic (Liu et al. 2004).
These three objects have low luminosity ratios, $L_{\rm IR}/L_* <2\times 10^{-2}$, and 
dust masses between $10^{-4}$ and $10^{-3}$\,M$_{\oplus}$, with an upper limit of 
about $0.5\,$M$_{\oplus}$. The collision time-scales for these three objects are 
extremely small (of the order of $100$\,yr). 

The preliminary models described above suggest that the disks around ROXs 36, 
RXJ1622.6-2345, IC348-56, and IC348-124 \emph{could} be younger analogs of 
the $\beta$ Pic and AU Mic debris disks, and thus some of the youngest debris 
disks ever observed. However, this interpretation depends on the assumption
that these four wTTs disks are gas poor (as their older counterparts are).
The timescale over which gas clearing occurs is still poorly constrained 
observationally and it is even unclear whether gas and dust are lost simultaneously
as disks evolve from the massive optically thick to the debris disk phase.
High-resolution \emph{Spitzer}-IRS observations, such as those
presented by Pascucci et al. (2007), to constrain the amount of gas 
present in these extremely young (1-3 Myrs) optically thin disks could provide crucial 
information on the gas evolution in the transition from the primordial to the 
debris disk stage.

In Section~\ref{DF}, we found that most wTTs ($\sim$80 $\%$) show no evidence for a disk. 
Since \emph{Spitzer} observations are capable of detecting very small amounts of dust in 
the planet-forming regions of the disk (r $\sim$0.1--10 AU), the absence of mid-IR excess imposes 
very stringent limits on the amount of dust available for planet formation around ``disk-less'' wTTs 
(i.e. most wTTs). In this section, we use the optically thin disk models discussed above 
to constrain the maximum amount of dust that could remain undetected within the 
first few tens of AU of the wTTs in our sample. Since the 24 $\mu$m  observations  are the most constraining, 
we restrict our analysis to Lupus and Ophiuchus objects that have 24 $\mu$m fluxes consistent with stellar photospheres 
and exclude the objects that remain undetected at this wavelength. Objects in IC 348 are also excluded because we 
detect the stellar photospheres of only a couple of them.  

Since none of our ``diskless''  wTTs are detected in the c2d 70 and 160 $\mu$m maps, for the 
purpose of our models, we adopt nominal upper limits of 15 and 250 mJy at 70 and 160 $\mu$m, respectively. 
For each model we calculated the mass encompassed within a radius $r$, as a function of this radius. 
With this approach, we can estimate the maximum dust mass in the inner regions of the wTTs with 
no detectable, or marginally detected, emission in excess to the photospheric emission.
The results are displayed in Figure \,\ref{MvsR} for the Lupus and Ophiuchus clouds, assuming 
distances to the Sun listed in Table 6. 
For the Lupus and Ophiuchus clouds, the c2d Spitzer observations constrain the observed wTTs 
to have less than a few $10^{-6}$\,M$_{\oplus}$ of dust within $1\,$AU from the central star, 
and less than a few $10^{-4}$\,M$_{\oplus}$ within $10\,$AU. These mass upper limits, obtained 
for minimum grain sizes between $10$ and $100\,\mu$m, drop by about an order of magnitude 
when $a_{\rm min} \leq 0.5\,\mu$m.The Lupus and Ophiuchus wTTs with no (or marginal) excess at $24\,\mu$m, have then inner
disks that are strongly depleted, and only extremely cold disks with
large inner holes are still theoretically possible because of
the relatively large MIPS $70\,\mu$m and $160\,\mu$m upper limits.
Such belts would resemble the dust rings resolved about nearby
young Main Sequence stars (e.g. HD\,181327, Schneider et al. 2006 and ref. therein).                                                             
Of course, our observations only constrain the mass of dust and can not rule out the presence 
of much larger planetesimals or planets because once dust grains grow into larger bodies 
(r $\gg$ $\lambda$), most of the solid mass never interacts with the radiation, and the opacity 
function, $k_{\nu}$ ($cm^2$/gr), decreases dramatically.

\subsection{Circumstellar Disks and Stellar Ages}\label{DF_vs_age} 

In Section~\ref{DF}, we studied the overall disk fraction of our sample of wTTs. In this section,
we derive stellar ages from two different evolutionary tracks and estimate the disk
frequency as a function of stellar age. In order to derive stellar ages from theoretical
evolutionary tracks, it is necessary to obtain the effective temperatures and luminosities of
all the targets. We estimate the effective temperatures directly from the spectral type of 
the objects according to the scale provided by Kenyon \& Hartmann (1995). We derive the stellar luminosities 
by applying a bolometric correction (appropriate for each spectral type) to the  2MASS J-band magnitudes 
corrected  for extinction and assuming the nominal cloud distance listed in Table 6. The J-band
was chosen because the effects of extinction is less important than at shorter wavelengths (A$_J$ $\sim$ 
0.26 A$_V$) and the emission from the disk is less prominent that at longer wavelengths. 
The bolometric corrections were taken from Hartigan, Strom \& Strom (1994) and the J-band extinction,  
$A_J$,  was calculated using 
$A_{J}= 1.24\times E(R_{C}-I_{C}) = 1.24\times ((R_{C}-I_{C})_{OBS}-(R_{C}-I_{C})_{O})$. 
The intrinsic stellar colors, (R$_{C}$-I$_{C}$)o, come from 
Kenyon \& Hartmann (1995). For objects without R$_{C}$ and I$_{C}$ fluxes available, we use 
$A_J$=5.88$\times$E(J-K). 

\subsubsection{Estimation of Age Uncertainties}

In order to estimate the error bars associated with the ages we derive, we first estimate
the observational uncertainties that need to be propagated through the H-R diagram. Fortunately,
the $T_{eff}$'s and luminosities of wTTs are easier to determine than those of cTTs. The interplay between 
extinction and veiling introduces a large uncertainty in determining stellar luminosities of cTTs, 
and different results are obtained depending on the band to which the bolometric correction is applied 
(Cieza et al. 2005). In fact, the luminosities obtained from the J-band are systematically 
larger (by $\sim$35$\%$) than those obtained from the $I_{C}$ band. For wTTs, Cieza et al. (2005) show that the luminosities 
derived from the $I_{C}$ and J bands agree to within 5$\%$; therefore, distance is probably the dominant  
uncertainty in calculating their luminosity. The distance uncertainties listed in Table 6 translate into 
$\sim$20--30$\%$ luminosity uncertainties. Similarly, the spectral type classification of wTTs (and therefore 
their temperatures) is usually more accurate than that of cTTs. This is because the spectra of cTTs are affected by 
veiling, and their photospheres are highly heterogeneous in terms of temperature due to the presence of hot accretion columns. 
To estimate the effective temperature uncertainty of the wTTs in our sample, we adopt one spectral type subclass as the 
classification accuracy.  For the reasons mentioned above, we regard the $T_{eff}$'s and the luminosities (and 
therefore the ages) of wTTs as being more accurate than those derived for cTTs using the same procedure.

A study of the disk fraction as a function of age can only yield meaningful results
if the intrinsic age spread of the sample is larger than the age uncertainties
attributable to observational errors. We verify that our sample satisfies this
condition by comparing the spread of the ages we derive to that expected solely
from observational errors. Based on an error budget similar to the one described above,
Hartman (2001) estimated that observational errors will introduce an age spread of
$\sigma$log(age) = 0.18 in the logarithmic age distribution derived for wTTs in Taurus.
In Figure 12, we  plot the derived age distribution of our sample as calculated from
two different evolutionary tracks. The mean age  and the age spread are model dependent.
According the the models  presented by D'Antona $\&$ Mazzitelli (1994, 1998, D98 hereafter)
\footnote{Available at http://www.mporzio.astro.it/~dantona/prems.html.}
the logarithmic age distribution of our sample can be characterized as a Gaussian
centered around log(age)=6.3 with $\sigma$(log age) = 0.57.  Similarly, according to the
models presented  by Siess et al. (2000, S00 hereafter)
\footnote{Available at http://www-astro.ulb.ac.be/~siess/database.html.}, 
the logarithmic age  distribution of the sample can be  characterized as a 
Gaussian centered around log(age)=6.6 with $\sigma$(log age) = 0.40. The age spreads 
derived are 2.2-3.2 times the value attributed by Hartmann (2001) to observational errors, 
which is also shown in Fig 13 for comparison.

\subsubsection{Evolutionary Tracks and Stellar Ages}

In order to evaluate the degree to which ages we derive depends on the models, 
we compare the individual ages derived from the D98 and S00 models. We choose these two particular evolutionary tracks 
because they provide the appropriate mass and age range and are 
both widely used, which allows a direct comparison of our results to those from other papers. 
Figure 13a shows the ages derived for our sample using both sets of evolutionary tracks. 
The error bars for every object have been calculated by propagating into the evolutionary tracks the 
\emph{observational} uncertainties computed as described in the previous section (e.g. a $T_{eff}$ uncertainty 
equal to one spectral type subclass and a luminosity uncertainty dominated by the distance uncertainty). Stellar  
ages of PMS stars are very difficult to estimate due to the large observational and model uncertainties involved 
(Hillenbrand $\&$ White, 2004) and they are often taken with a high degree of (healthy) skepticism. 
However, even though the the  error bars in the individual ages are large, the total age spread in the sample is 
significantly larger than the typical (observational) error bar. This is consistent with 
the analysis of Figure 12. Also, even though D98 and S00 evolutionary tracks show some systematic differences 
(e.g. D98 tracks yield significantly younger ages than S00 models), the relative ages agree fairly well. 

Figure 13a shows that $\sim$40$\%$ of the wTTs that are both younger than 1.5 Myrs according to the S00 models 
\emph{and} younger than 0.6 Myrs according to the D98 models have circumstellar disks. In contrast, none of the 
targets that are older than 10 Myrs according to the D98 \emph{or} S00 models has a disk. 
The decrease in the IRAC disk fraction with stellar age is clearly seen Figure 13b, where we 
restrict our analysis to the 198 objects in the magnitude limited subsample discussed in 
Section ~\ref{DF_stats} and divide the ages we derive using the D98 and S00 models into 4 age bins.  
Very similar conclusions can be drawn if our sample is combined with that of P06. Including the P06 
samples considerably increases the statistical significance of the last age bin. Taken both samples together, 
none of the $\sim$40 stars that are older than 10 Myrs according to either of the tracks have an 
IR excess indicating the presence of a circumstellar disk.
                                                                                                             
\subsection{Constraint on the Timescale of Planet Building}\label{constrains}

Figure 13 suggests that circumstellar disks, as defined by the presence of IR excesses at 
$\lambda$ $\leq$ 10-24 $\mu$m, are very rare or nonexistent around wTTs with ages $\gtrsim$10 Myrs.
This timescale  is very similar to that obtained by studies of the frequency of circumstellar disks detected 
in the near-IR (See Hillenbrand  2006 for a review). However, our results impose much stronger constraints on 
the time available for the formation of planets than those provided by previous studies. Past results based on near-IR 
excesses always left room for the possibility that stars without near-IR excess had enough material to form planets at larger 
radii not probed by near-IR  wavelengths. IRAS and ISO had the appropriate wavelengths range to probe the planet-forming 
regions of the disk but lacked the sensitivity needed to detect all but the strongest  mid- and far-IR excesses in 
low-mass stars at the distances of nearest star-forming regions. \emph{Spitzer} provides, for the first time, the 
wavelength coverage and the sensitivity needed to detect small amounts of dust in the planet-forming regions of 
a statistically significant number of low-mass PMS stars. In particular, the results from Section~\ref{DISK_MODELS} 
suggest that 24 $\mu$m fluxes consistent with stellar photospheres constrain the amount of \emph{warm} 
dust (T $\sim$100 K) in the disks of our sample of wTTs to be much less than an Earth mass. 
Even though our 24 $\mu$m observations are not sensitive to the stellar photospheres 
of all the targets, taking the P06 sample and our sample together, there are at 
least $\sim$40 wTTs with estimated ages $>$ 10 Myrs showing photospheric fluxes 
at 24 $\mu$m. This number is likely to be a lower limit because, as discussed 
in Section \ref{L06_comp}, we suspect that the ages of many of the stars in the P06 sample 
have been underestimated. This seems to imply that after $\sim$10 Myrs wTTs have 
to be in a relatively advanced stage of the planet formation process if they are to 
form planets at all. 

Since the cTTs disks older than 10 Myrs are also very infrequent, 10 Myrs seems to be a general
upper limit for the survival of primordial disks around PMS stars. This conclusion
is also supported by recent results from the ``Formation and Evolution of Planetary Systems'' (FEPS)
project. Silverstone et al. (2006, S06 hereafter) search for IRAC excesses in a sample of 74 young
(age $<$ 30 Myr old) Sun-like (0.7 $<$ M$_*$/M$_{\odot}$) stars. They divided
the sample into two age bins, 3-10 Myrs and 10-30 Myrs.  S06 find IRAC excesses for 4 of the
29 stars in the youngest age bin and for 1 of the 45 stars in the older age bin.  The FEPS
objects were selected based on their ages ($<$ 30 Myrs), masses ($\sim$ 1 M$_{\odot}$), distances
($<$ 170 pc), and low infrared backgrounds, without a bias with respect to their H$\alpha$ EW
or IR properties.  The five objects with IRAC excess have SEDs consistent with those of CTTS, and
given the age uncertainties, it entirely possible that they are all younger 
than 10 Myrs. In fact, the only object with IRAC excess for which S06 adopts an age older than 
10 Myrs is PDS 66. S06 adopted an age of 17 Myrs for PDS 66 based on the mean age of 
the Lower Centaurus Crux,  but its formal age is 7-17 Myrs depending on the evolutionary track 
used (Mamajek et al. 2002). Recent sub-millimeter results extend the conclusions on the survival time 
of the material in the inner disk (r $<$ 0.1 AU) and the planet-forming region of the disk (r $\sim$ 1--10 AU)
to the outer disk (r $\sim$100 AU). Andrews $\&$ Williams (2005) observed over 150 YSOs in Taurus and 
found that $<$ 10 $\%$ of the object lacking inner disk signatures were detected at sub-mm wavelengths. 
Given the high sensitivity of their survey (3$\sigma$ $\sim$10 mJy at 850 $\mu$m) they conclude that dust
in the inner and outer disk dissipate nearly simultaneously. 
                                                                                                                
Figures 13 also suggests that the disks of some wTTs dissipate in timescales $\lesssim$1 Myrs\footnote{This assumes 
that all wTTs had a disk at some point of their evolution. It is possible that some disks dissipate very early 
(t $<<$  1 Myrs), even before the star is optically revealed and can be classified as a wTTs. However, accretion 
through a disk is considered an unavoidable step of the star formation process.}. Are these apparently very young disk-less 
objects as young as the evolutionary tracks suggest or is their apparent youth just a product of the large age uncertainty?  
Several factors can introduce very large errors in the age determination. For instance, the luminosity of foreground field stars 
can be grossly overestimated if the nominal cloud distances are used, leading to grossly underestimated ages. Also, large errors 
in the spectral type classification might lead to large errors in the extinction and luminosities.  
To check for these possibilities,  we plotted the SEDs of the objects classified as the youngest disk-less wTTs in our 
sample. Their SEDs look consistent with stellar photospheres and the overall quality of the fits suggests that both the 
spectral types and the extinction corrections are reasonably accurate. Still, some of these objects are $\sim$0.5 Myrs 
old according to the DM98 models and $\sim$1.0 Myrs old according to the S00 models, and $\sim$15 times more luminous than 
main-sequence stars of the corresponding spectral types. 

In Section~\ref{DF_vs_age} we found that the  dissipation timescale or ``survival time'' of 
wTTs circumstellar disks ranges from  less than 1 to 10 Myrs. A related timescale is the transition 
timescale from optically thick accretion disks to undetectable disks. Assuming that wTTs disks are the 
link between cTTs disks and ``disk-less'' wTTs, then the transition timescale, $\tau$, can be estimated as 
$\tau$= $\frac{N_{TRAN}}{N_{PMS}}\times<$age$>$, where $N_{TRAN}$ is the number of wTTs disks, 
$N_{PMS}$ is the total number of PMS stars (wTTs+cTTs), and $<$age$>$ is the mean age of the sample. 
Adopting the wTTs/cTTs ratio of IC 348 ($\sim$3/2) and a mean age of 3 Myrs, the overall IRAC disk fraction 
in wTTs of 20$\%$ we find in Section~\ref{DF} implies $\tau\sim$0.4 Myrs. This timescale is very similar 
to that found for the transition timescale between an optically thick disk and an optically thin disk 
by Skrutskie et al. (1990) and Wolk $\&$ Walter (1996) based on the number of ``transition objects'', 
which they define as targets without K-band excess but with strong IRAS excesses. The fact that the 
transitional timescale is significantly shorter than the mean disk-lifetime is inconsistent with traditional 
viscous evolution (Hartmann et al. 1998) or magnetospheric clearing models (Armitage et al. 1999), which 
predict a steady disk evolution and thus similar timescales for disk lifetimes and disk dispersal times.
Non-steady disk evolution scenarios are required to explain the short transitional timescales inferred  
after significantly longer disk lifetimes. Such scenarios include the ultraviolet-switch model 
(Alexander et al. 2006) and the presence of gap forming planets (e.g. Quillen et al. 2004).  

Our results from Section 4.4, constrain not only the dissipation timescale of the dust during the planet 
formation process, but also the amount of second generation dust that is produced during this process 
(Section~\ref{DISK_MODELS}). Numerical models presented by Kenyon 
$\&$ Bromley (2004) predict the amount of 10 and 20 $\mu$m excess as a function of time produced by the 
formation of terrestrial planets. Detailed comparison between these kind of models with predictive power and 
the new observational constraints \emph{Spitzer} is now providing could be highly valuable for our 
understanding of planet formation.  

Particularly intriguing and potentially very important objects are the $\sim$1 Myrs old wTTs without any measurable 
IR excess (for $\lambda$ $\le$ 24 $\mu$m) discussed above. One possible explanation for the very early dissipation of 
their disks is that these stars have already formed planets through gravitational instability, which is expected to 
occur at extremely young ages when disks are most massive (Boss 2000). Another possibility for the fast dissipation 
of these disks is the presence of close companions that could have disrupted their disks. This possibility can be 
tested with a combination of radial velocity and adaptive optics observations to search for companions. Initial 
conditions could also be responsible for the early dissipation of the disk, although this hypothesis is not 
easily testable observationally. We note that the Space Interferometry Mission should be able to establish 
whether or not these very young disk-less wTTs do in fact harbor planets. The presence of planets around these 
$\sim$1 Myrs old stars would set the tightest constraints to date for the planet formation timescale.

The properties of wTTs with disks such as their age, SED morphology, $L_{DISK}/L_{*}$, etc, strongly 
suggests that wTTs disks are the link between the massive primordial disks found around cTTs and the debris disks observed 
around young MS stars. They could be arguably the best places to study ongoing planet formation. Before the 
end of its mission, \emph{Spitzer} is likely to identify hundreds of wTTs disks in nearby star-forming 
regions. These objects will most likely be the  main targets of many  follow-up observations. Deep far-IR and 
sub-mm observations with \emph{Spitzer}, Herschel, and Alma will allow the study of the outer regions of 
these wTTs disks and estimates of their masses. Follow-up \emph{Spitzer}/IRS observations of the 10 and 20 $\mu$m 
silicate features will provide important information on the evolutionary state of the circumstellar dust around 
these objects. Finally, high resolution searches for $H_{2}$ and atomic lines, such as
those presented by Hollenbach et al. (2005) and Pascucci et al. (2007), would be highly desirable to constrain 
the amount of gas available for the formation of giant planets in wTTs disks. 

\section{Summary of results}\label{sumary}

We present a census of circumstellar disks and report the disk frequency as a function of stellar age for a sample 
of over 230 spectroscopically identified wTTs located in the c2d IRAC and MIPS maps of the Ophiuchus, Lupus 
and Perseus Molecular Clouds. Our main results can be summarized as follows:

1) In Section~\ref{DF}, we find from a magnitude limited subsample of wTTs that $\sim$20$\%$ of the wTTs 
have noticeable IR-excesses at IRAC wavelengths indicating  the presence of a circumstellar disk. 

2) The disks frequencies we find in the 3 clouds we consider are $\sim$3-6 times larger than that recently found by P06 for a 
sample of 83 relatively isolated wTTs projected outside the boundaries of nearby molecular clouds. This discrepancy in the 
disk fractions  supports the idea that samples of wTTs (nominally 1-10 Myrs old) located a few degrees away from their 
parent molecular clouds represent an older population of stars. The disk fractions we find are more consistent with those 
obtained in recent \emph{Spitzer} studies of wTTs in young clusters such as IC 348 and Tr 37. 
However, in Section~\ref{L06_comp}, 
we suggest that Lada and colleagues might have overestimated the disk fraction of wTTs in IC 348 by classifying as 
``anemic disks'' some disk-less M4-M7 stars with large photometric uncertainties.

3) In Section~\ref{DF_vs_Halpha}. we find that the disk fraction of wTTs is a smooth function 
of $H\alpha$ EW. In Section~\ref{FDL}, we show that the fractional disk luminosities of wTTs 
disks bridge the gap between the cTTs and the debris disk range.

4) In Section~\ref{DISK_MODELS}, we estimate mass upper limits of dust within the inner 10 AU 
of $10^{-4} M_{\oplus}$ for the objects in our sample with 24 $\mu$m fluxes
consistent with stellar photospheres. 

5) In Section~\ref{DF_vs_age}, we place our sample of wTTs in the H-R diagram and find that the stars with excesses are among the 
younger part of the age distribution. However, we also find that up to $\sim$50$\%$ of the apparently  youngest stars  
in the wTTs sample show no evidence of IR excess. This suggests that the circumstellar disks of a sizable fraction of 
pre-main-sequence stars dissipate before the stars reach an age of $\sim$1 Myr. 

6) Also in Section~\ref{DF_vs_age}, we find that none of the stars in our sample apparently older than $\sim$10 Myrs have 
detectable circumstellar disks. Since \emph{Spitzer}  observations probe planet-forming regions of the disk (r $\sim$0.1-10 AU) 
and are capable of detecting IR excesses produced by very small amounts of dust, our results impose stronger 
constraints on the time available for the formation of planets than those provided by previous studies based on 
detections of disks in the near-IR.

7) Finally, in Section~\ref{constrains} we estimate a transition timescale of $\sim$0.4  Myrs 
between  optically thick accretion disks and disks that are undetectable shortward of $\sim$10 $\mu$m,
in good agreement with previous results. 

\acknowledgments

We thank the anonymous referee for his/her many detailed comments, 
which have helped to improve the paper.
Support for this work, which is part of the {\it Spitzer} Legacy
Science Program, was provided by NASA through contracts 1224608,
1230782, and 1230799 issued by the Jet Propulsion Laboratory,
California Institute of Technology under NASA contract 1407. 
B.M. acknowledges the Fundaci\'on Ram\'on Areces for financial support.
Astrochemistry in Leiden is supported by a NWO Spinoza and
NOVA grant,  and by the European Research Training Network
"The Origin of Planetary Systems" (PLANETS, contract number
HPRN-CT-2002-00308). We thank the Lorentz Center in Leiden for 
hosting several meetings that contributed to this paper.
This publication makes use of data products from the Two Micron All Sky
Survey, which is a joint project of the University of Massachusetts
and the Infrared Processing and Analysis Center funded by
NASA and the National Science Foundation. We also acknowledge
use of the SIMBAD database.

\clearpage

\tablenum{1}
\thispagestyle{empty}
\begin{deluxetable}{lcccccccccccc}
\rotate
\tabletypesize{\footnotesize}
\tiny
\tablewidth{0pt}
\tablecaption{\emph{Spitzer data}}
\tablehead{\colhead{ID}&\colhead{R.A}&\colhead{Dec}&\colhead{F$_{3.6}$}&\colhead{Error$_{3.6}$}&\colhead{F$_{4.5}$}&\colhead{Error$_{4.5}$}&\colhead{F$_{5.8}$}&\colhead{Error$_{5.8}$}&\colhead{F$_{8.0}$}&\colhead{Error$_{8.0}$}&\colhead{F$_{24}$}&\colhead{Error$_{24}$}\\
\colhead{}&\colhead{(J2000.0)}&\colhead{(J2000.0)}&\multicolumn{10}{c}{(mJy)}}
\startdata
IC348-1 & 55.8837 & 32.1048 & 1.06e+01 & 1.51e-01 & 7.09e+00 & 8.78e-02  & 4.81e+00 & 5.52e-02 & 2.79e+01 & 4.68e-02 & --- & ---\\
IC348-2 & 55.8903 & 32.0293 & 4.96e+00 & 7.22e-02 & 3.42e+01 & 4.89e-02  & 2.52e+00 & 3.64e-02 & 1.42e+01 & 2.89e-02 & --- & ---\\
IC348-3 & 55.9526 & 32.2308 & 3.23e+00 & 4.64e-02 & 2.39e+00 & 2.82e-02  & 1.65e+00 & 2.96e-02 & 9.38e-01 & 3.05e-02 & --- & ---\\
IC348-4 & 55.9532 & 32.1259 & 9.01e+00 & 1.75e-01 & 6.51e+00 & 8.82e-02  & 4.13e+00 & 5.04e-02 & 2.34e+00 & 4.26e-02 & --- & ---\\
IC348-5 & 55.9558 & 32.1778 & 7.47e+00 & 1.13e-01 & 5.55e+00 & 6.76e-02  & 3.57e+00 & 4.47e-02 & 2.13e+00 & 4.04e-02 & --- & ---\\
\enddata
\tablecomments{[The complete version of this table is in the electronic edition of the Journal. The printed edition contains only a sample
to illustrate its content.]}
\end{deluxetable}

\clearpage

\tablenum{2}
\thispagestyle{empty}
\begin{deluxetable}{lccccccccccccc}
\tabletypesize{\footnotesize}
\rotate
\tiny
\tablewidth{0pt}
\tablecaption{non-\emph{Spitzer data}}
\tablehead{\colhead{ID}&\colhead{R.A}&\colhead{Dec}&\colhead{SpT}&\colhead{H$_{\alpha}$}&\colhead{V}&\colhead{Error$_V$}&\colhead{R$_C$}&\colhead{Error$_{R_{C}}$}&\colhead{I$_C$}&\colhead{Error$_{I_{C}}$}&\colhead{J}&\colhead{$H$}&\colhead{K$_{S}$}\\
\colhead{}&\colhead{(J2000.0)}&\colhead{(J2000.0)}&\colhead{}&\colhead{($\AA$)}&\multicolumn{9}{c}{(mag)}}
\startdata
IC348-1 & 55.8837 & 32.1048 & M0.75 & 1.0 & --- & ---   & 14.81 & 0.03  & 13.73 & 0.03 & 12.18 & 11.38 & 11.13 \\
IC348-2 & 55.8903 & 32.0293 & M5    & 5.0 & --- & ---   & 17.93 & 0.03  & 15.91 & 0.03 & 13.36 & 12.57 & 12.22 \\
IC348-3 & 55.9526 & 32.2308 & M5    &   6 & --- & ---   & 17.92 & 0.03  & 16.05 & 0.03 & 13.59 & 12.98 & 12.64 \\
IC348-4 & 55.9532 & 32.1259 & M1.5  &   0 & --- & ---   & 15.44 & 0.03  & 14.16 & 0.03 & 12.44 & 11.59 & 11.34 \\
IC348-5 & 55.9558 & 32.1778 & M3.5  &  11 & --- & ---   & 16.36 & 0.03  & 14.76 & 0.03 & 12.58 & 11.80 & 11.54 \\
\enddata
\tablecomments{[The complete version of this table is in the electronic edition of the Journal. The printed edition contains only a sample
to illustrate its content.]}
\end{deluxetable}

\tablenum{3}
\thispagestyle{empty}
\begin{deluxetable}{lcc}
\tablewidth{0pt}
\tablecaption{Adopted Extinction Relations}
\tablehead{\colhead{    }&\colhead{$\lambda$}&\colhead{$A_{V}$/$A_{\lambda}$\tablenotemark{1}}\\
           \colhead{Band}&\colhead{($\mu$m) }&\colhead{}}
\startdata
V         & 0.55  & 1.00 \\
R$_{C}$   & 0.65  & 0.79 \\
I$_{C}$   & 0.80  & 0.58 \\
J         & 1.25  & 0.26 \\
H         & 1.66  & 0.15 \\
K$_{S}$   & 2.2   & 0.09 \\
IRAC-1    & 3.6   & 0.04 \\
IRAC-2    & 4.5   & 0.03 \\
IRAC-3    & 5.8   & 0.02 \\
IRAC-4    & 8.0   & 0.01 \\
MIPS-1    & 24    & 0.002 \\
\enddata
\tablenotetext{1}{The extinction relations for the optical and 2MASS wavelengths come 
the Asiago database of photometric systems (http://ulisse.pd.astro.it/Astro/ADPS/enter.html ; 
Fiorucci \& Munari 2003), while those for the \emph{Spitzer} bands come from Huard et al. 
(2007, in prep.}
\end{deluxetable}

\clearpage

\tablenum{4}
\thispagestyle{empty}
\begin{deluxetable}{lccccccccc}
\rotate
\tabletypesize{\footnotesize}
\tablewidth{0pt}
\tablecaption{Objects classified as anemic disks by L06 that do not satisfy our disk identification criterion}
\tablehead{\colhead{Ra   }&\colhead{Dec}&\colhead{ID}&\colhead{SpT}&\colhead{$\alpha$}&\colhead{$\sigma$$_{\alpha}$}&\colhead{($\alpha$+2.66)/$\sigma$$_{\alpha}$}&\colhead{[3.6]-[8.0]}&\colhead{[3.6]-[8.0]}&\colhead{$\Delta$([3.6]-[8.0])}\\
\colhead{(J2000.0)}&\colhead{(J2000.0)}&\colhead{L06}&\colhead{}&\colhead{L06}&\colhead{L06}&\colhead{L06}&\colhead{L06}&\colhead{C2D}&\colhead{L06-C2D}}
\startdata
55.9526 & 32.2308 &         261 & M5    &  -2.560 & 0.048 & 2.0 &  0.29  &          0.26  &         0.03 \\
55.9742 & 32.1251 &         254 & M4.25 &  -2.549 & 0.081 & 1.3 &  0.30  &          0.29  &         0.01 \\
56.0185 & 32.0817 &         303 & M5.75 &  -2.197 & 0.125 & 3.7 &  0.61  &          ---   &         ---  \\
56.0740 & 32.0799 &         169 & M5.25 &  -2.440 & 0.099 & 2.2 &  0.41  &          0.36  &         0.05 \\
56.0816 & 32.0403 &         322 & M4.25 &  -2.426 & 0.046 & 5.0 &  ---   &          ---   &         ---  \\
56.0971 & 32.0318 &        1684 & M5.75 &  -1.811 & 0.500 & 1.6 &  ---   &          ---   &         ---  \\
56.1203 & 32.0730 &         385 & M5.75 &  -2.183 & 0.350 & 1.3 &  0.71  &          ---   &         ---  \\
56.1309 & 32.1915 &        226  & M5.25 &  -2.233 & 0.265 & 1.6 &  0.60  &          0.38  &         0.22 \\
56.1358 & 32.1451 &          33 & M2.5  &  -2.483 & 0.157 & 1.1 &  0.32  &          0.16  &         0.16 \\
56.1365 & 32.1544 &          88 & M3.25 &  -2.355 & 0.081 & 3.7 &  0.46  &          0.48  &        -0.02 \\
56.1460 & 32.1269 &        8024 & K7    &  -2.213 & 0.218 & 2.0 &  0.62  &          0.25  &         0.37 \\
56.1590 & 32.1727 &         353 & M6    &  -2.251 & 0.178 & 2.2 &  0.61  &          0.16  &         0.45 \\
56.1794 & 32.1709 &         217 & M5    &  -2.312 & 0.043 & 8.0 &  0.52  &          0.43  &         0.09 \\
56.1822 & 32.1800 &         360 & M4.75 &  -2.303 & 0.064 & 5.5 &   ---  &          ----  &         ---  \\
56.1861 & 32.1251 &         218 & M5    &  -2.294 & 0.049 & 7.4 &  0.54  &         -0.04  &         0.58 \\
56.1902 & 32.1864 &         413 & M4    &  -2.398 & 0.500 & 0.5 &  ---   &          0.31  &         ---  \\
56.2035 & 32.2228 &         178 & M2.75 &  -2.520 & 0.080 & 1.7 &  0.31  &          0.23  &         0.08 \\
56.2527 & 32.1387 &         344 & M5    &  -2.376 & 0.229 & 1.2 &  0.49  &          0.45  &         0.04 \\
\enddata
\end{deluxetable}

\clearpage

\tablenum{5}
\thispagestyle{empty}
\begin{deluxetable}{lcccccccc}
\tabletypesize{\footnotesize}
\rotate
\tablewidth{0pt}
\tablecaption{Optically thin disk properties, for two different regimes of
minimal grain sizes, $a_{\rm min}$. }
\tablehead{\colhead{}&\multicolumn{3}{c}{(a$_{min}$ $<$ 0.5 $\mu$m)}&\multicolumn{3}{c}{(a$_{min}$ $>$ 10 $\mu$m)}&\colhead{}\\
\colhead{Star}&\colhead{r$_{0}$(AU)}&\colhead{M$_{dust}$(10$^{-3}$M$_{\oplus}$)}&\colhead{t$_{coll}$(yr)}&\colhead{r$_{0}$(AU)}&\colhead{M$_{dust}$(10$^{-3}$M$_{\oplus}$)}&\colhead{t$_{coll}$(yr)}&\colhead{L$_{IR}$/L$_{*}$$\times$10$^{3}$}}
\startdata
ROXs 36 &
$     82_{-     73}^{+ 100} $ &
$     6.7_{-     6.5}^{+     25} $ &
$ 77000_{- 74000}^{+ 110000} $ &
$     17_{-     10}^{+     2.8} $ &
$     5.3_{-     4.9}^{+     5.9} $ &
$ 7400_{- 5200}^{+ 1900} $ &
$    0.11_{-   0.023}^{+   0.015} $ \\
RXJ1622.6-2345 &
$     4.0_{-     3.5}^{+     51} $ &
$    0.13_{-    0.13}^{+ 280} $ &
$ 310_{- 300}^{+ 2000} $ &
$     1.9_{-     1.4}^{+     6.1} $ &
$    0.35_{-    0.34}^{+     85} $ &
$ 130_{- 110}^{+     79} $ &
$    0.60_{-    0.11}^{+     5.4} $ \\
IC348-56 &
$     5.2_{-     4.4}^{+     59} $ &
$    0.53_{-    0.52}^{+ 570} $ &
$ 190_{- 190}^{+ 1300} $ &
$     2.5_{-     1.8}^{+     6.7} $ &
$     1.5_{-     1.5}^{+ 180} $ &
$     86_{-     70}^{+     52} $ &
$     1.4_{-    0.27}^{+     9.7} $ \\
IC348-124 &
$     4.1_{-     3.2}^{+     27.} $ &
$    0.74_{-    0.69}^{+ 170} $ &
$     65_{-     61}^{+ 330} $ &
$     1.8_{-    0.94}^{+     2.4} $ &
$     1.8_{-     1.6}^{+     33} $ &
$     24_{-     14}^{+     15} $ &
$     3.1_{-    0.65}^{+     8.6} $\\
\enddata
\end{deluxetable}

\clearpage

\tablenum{6}
\thispagestyle{empty}
\begin{deluxetable}{lcc}
\tablewidth{0pt}
\tablecaption{Adopted Distances}
\tablehead{\colhead{Cloud}&\colhead{Distance}&\colhead{Reference}\\
           \colhead{}     &\colhead{(pc)}    &\colhead{}}
\startdata
Ophiuchus         & 125$\pm$20 &  de Geus et al.    (1989)          \\
Lupus I           & 150$\pm$20 &  Comeron et al. (2006), in prep.\\  
Lupus III         & 200$\pm$20 &  Comeron et al. (2006), in prep.\\
IC348             & 320$\pm$30 &  Herbig (1998)                  \\
\enddata
\end{deluxetable}

\clearpage

\begin{figure}
\figurenum{1a and 1b}
\plottwo{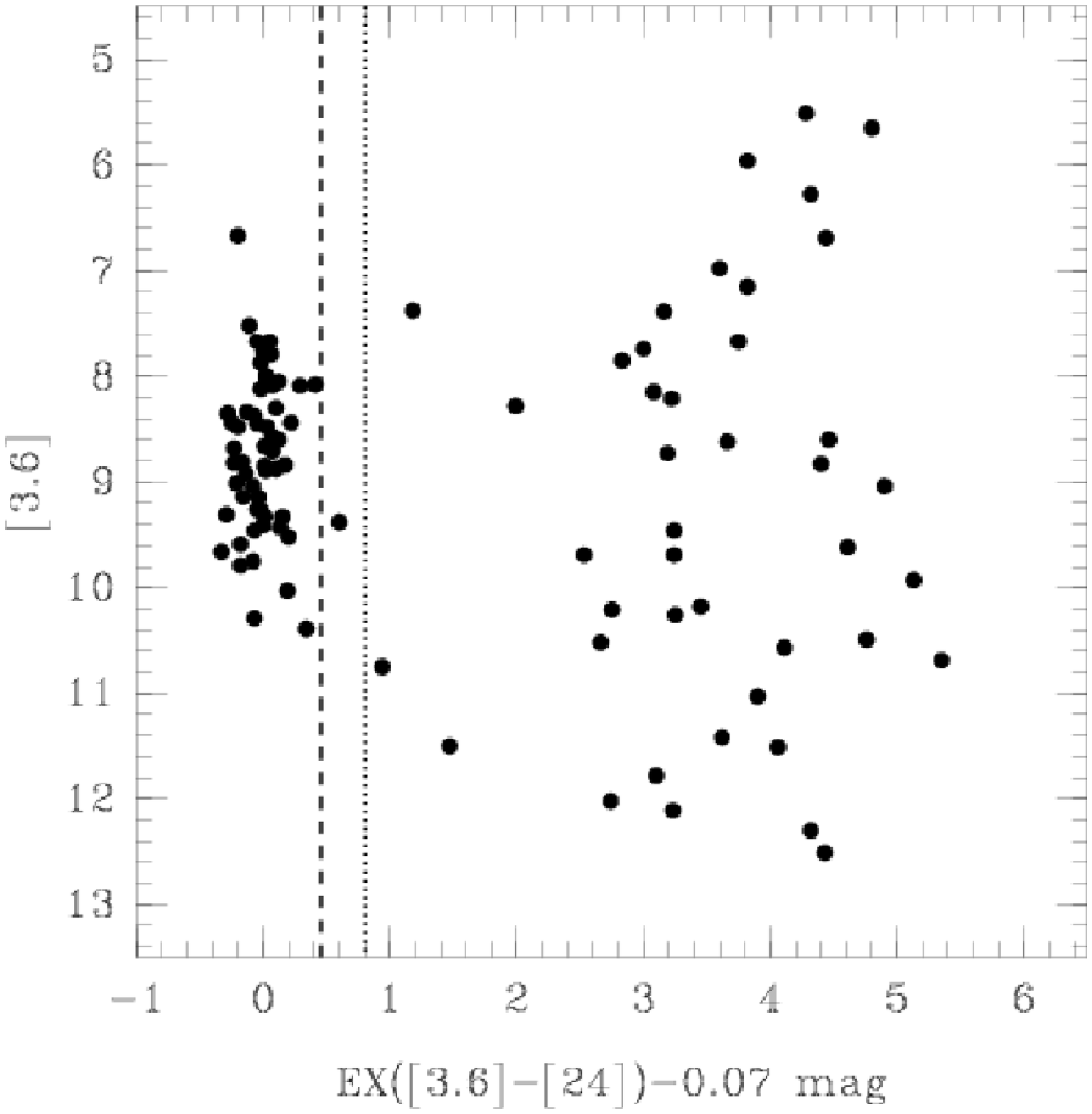}{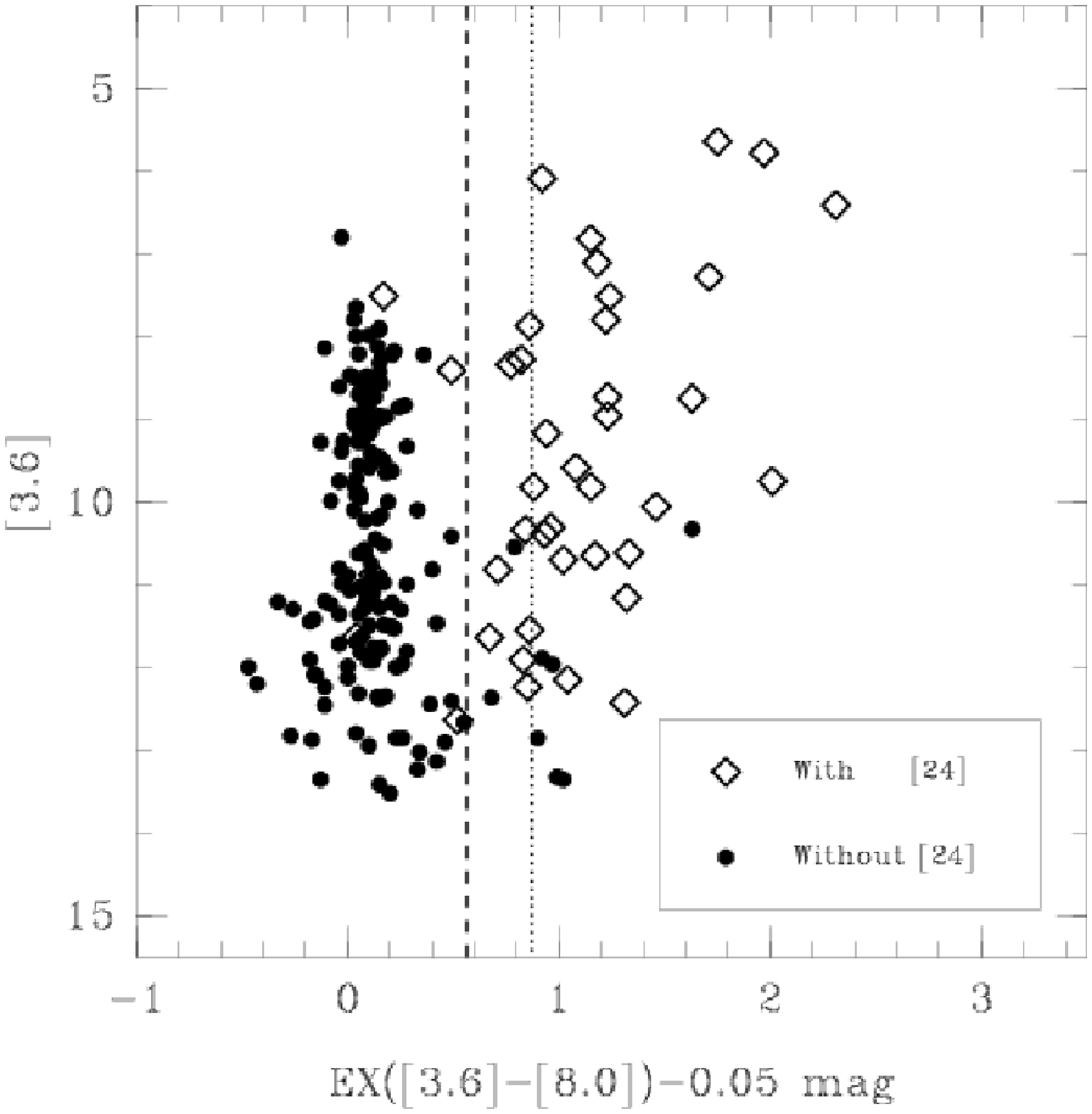} 
\caption{[3.6] vs EX([3.6]--[24]) (left) and  [3.6] vs EX([3.6]--[8.0]) (right) diagrams for 
our sample of wTTs used for disk identification. See text for definitions. 
The 3 and 5-$\sigma$ dispersion of the 
stellar photospheres are shown as dashed and dotted lines, respectively. Most of the 
disks are detected at 24 $\mu$m.}
\end{figure}

\begin{figure}
\epsscale{1}
\figurenum{2}
\plotone{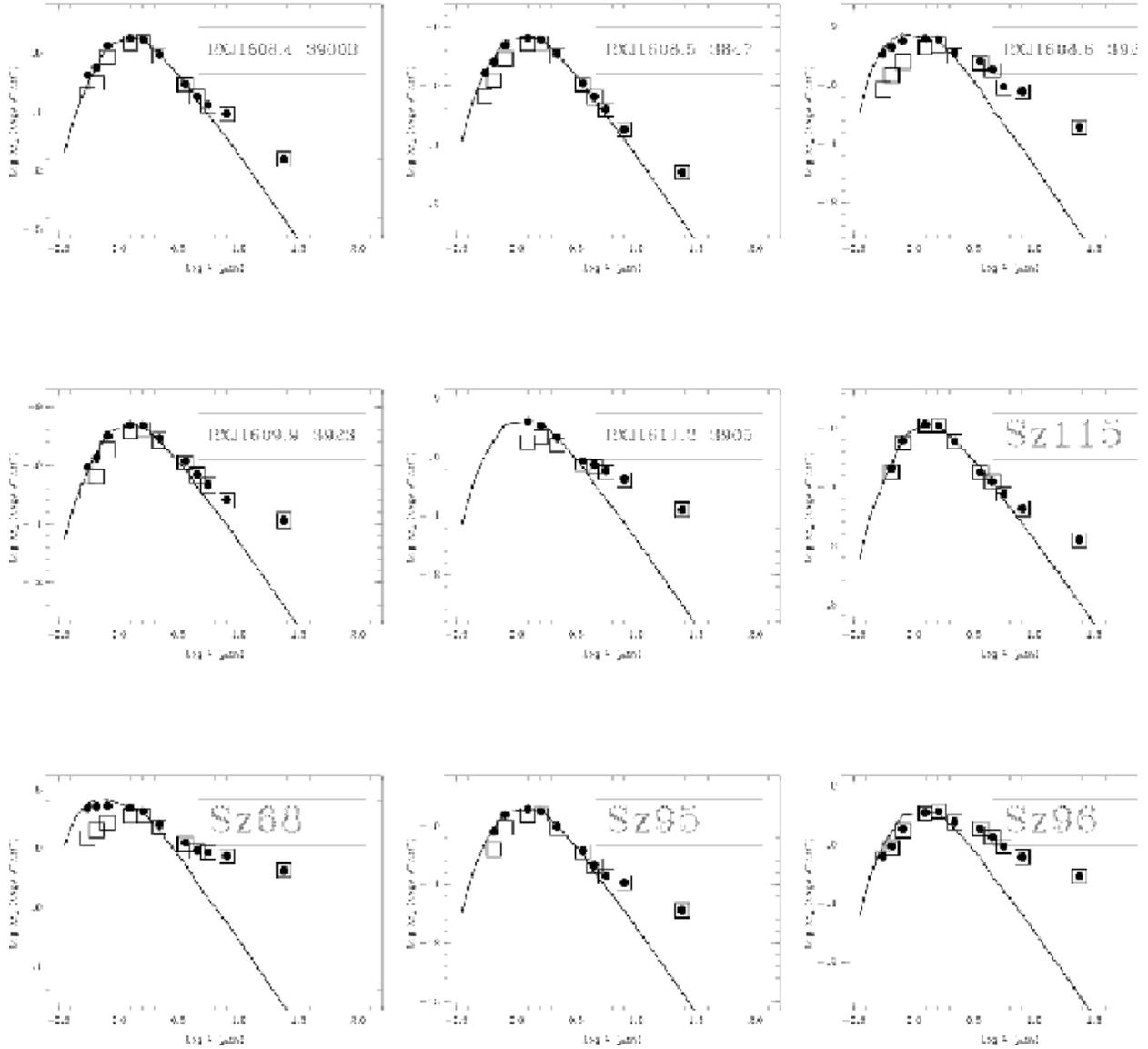}
\caption{SEDs of wTTs disks in Lupus.
The open squares represent the observed optical, 2MASS, IRAC 
and MIPS-24 $\mu$m fluxes, while the dots correspond to the 
extinction corrected values. Model photospheres corresponding 
to published spectral types are shown for comparison.}
\end{figure}

\begin{figure}
\epsscale{1}
\figurenum{3}
\plotone{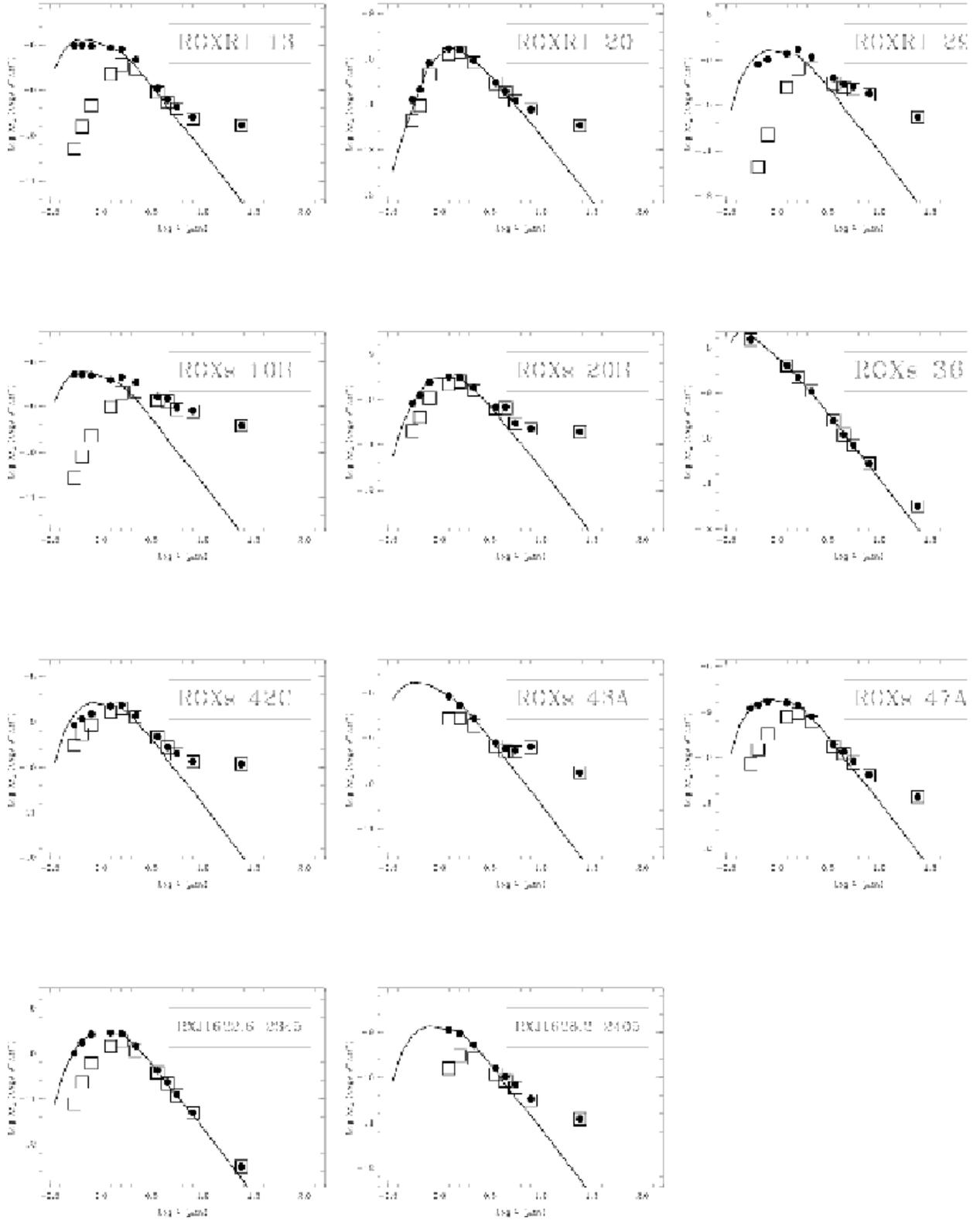}
\caption{SEDs of wTTs disks in Ophiuchus.}
\end{figure}
\clearpage
\begin{figure}
\epsscale{.9}
\figurenum{4}
\plotone{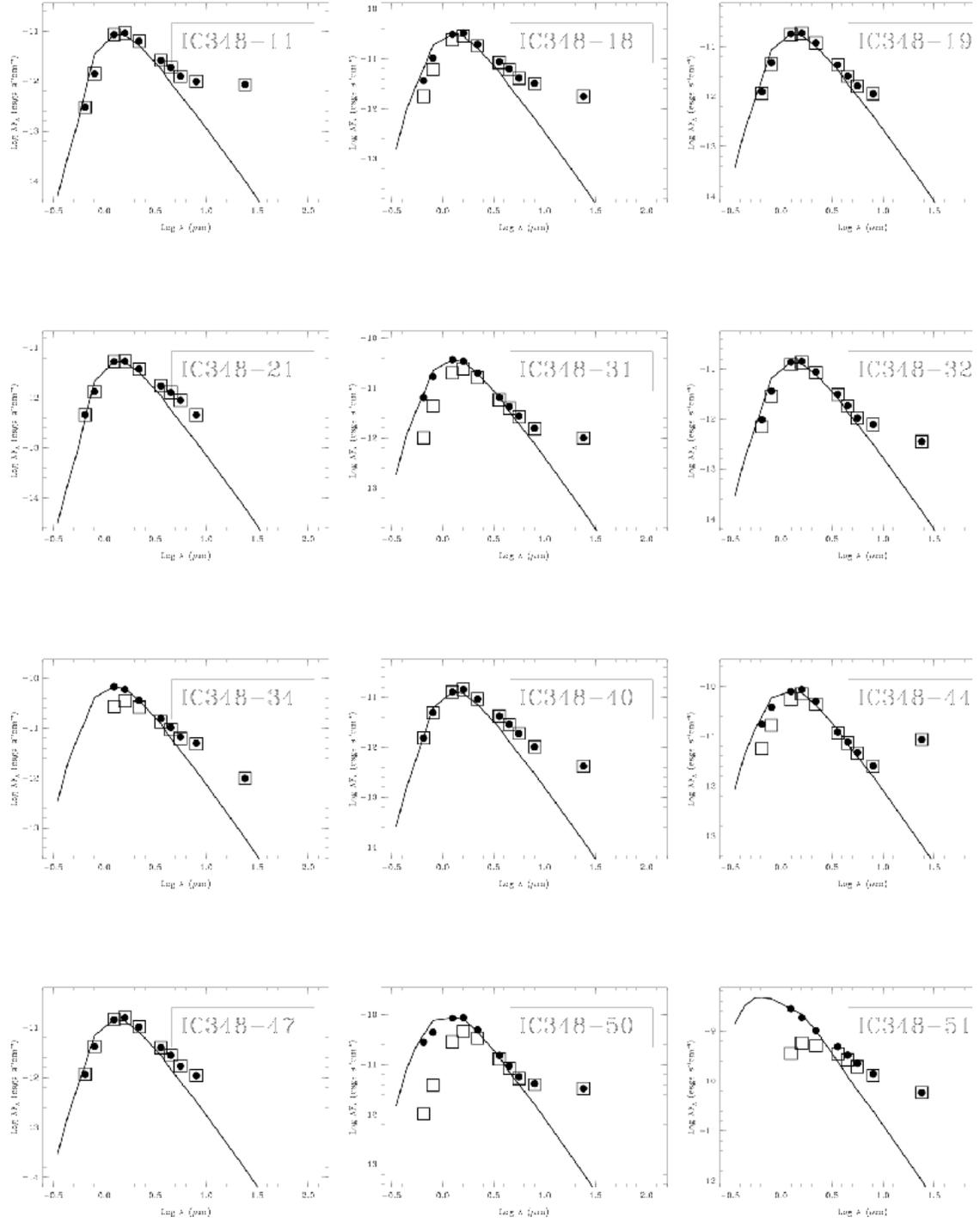}
\caption{SEDs of wTTs disks in IC 348.}
\end{figure}
\clearpage
{\plotone{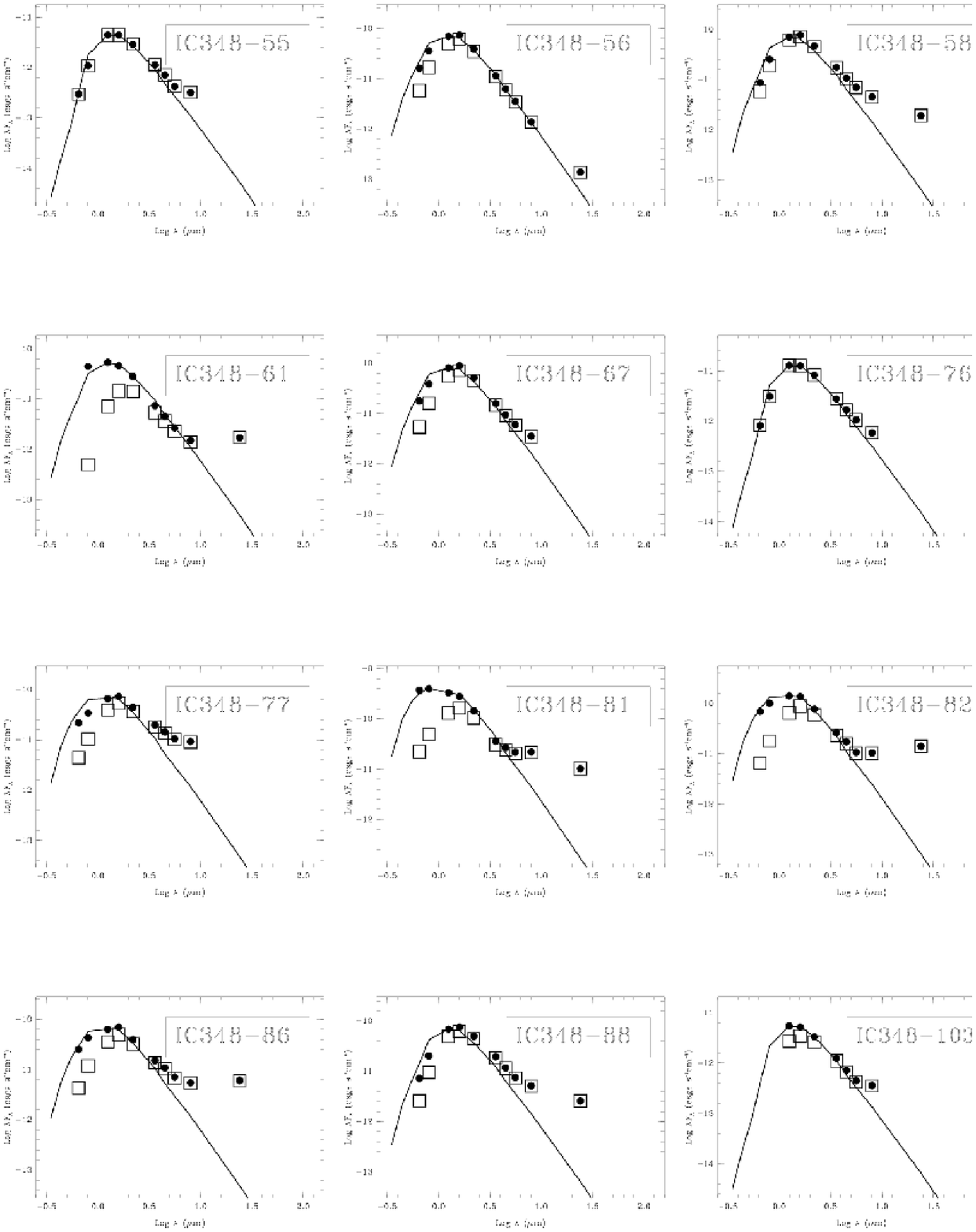}}\\[5mm]
\centerline{Fig. 4. --- Continued.}
\clearpage
{\plotone{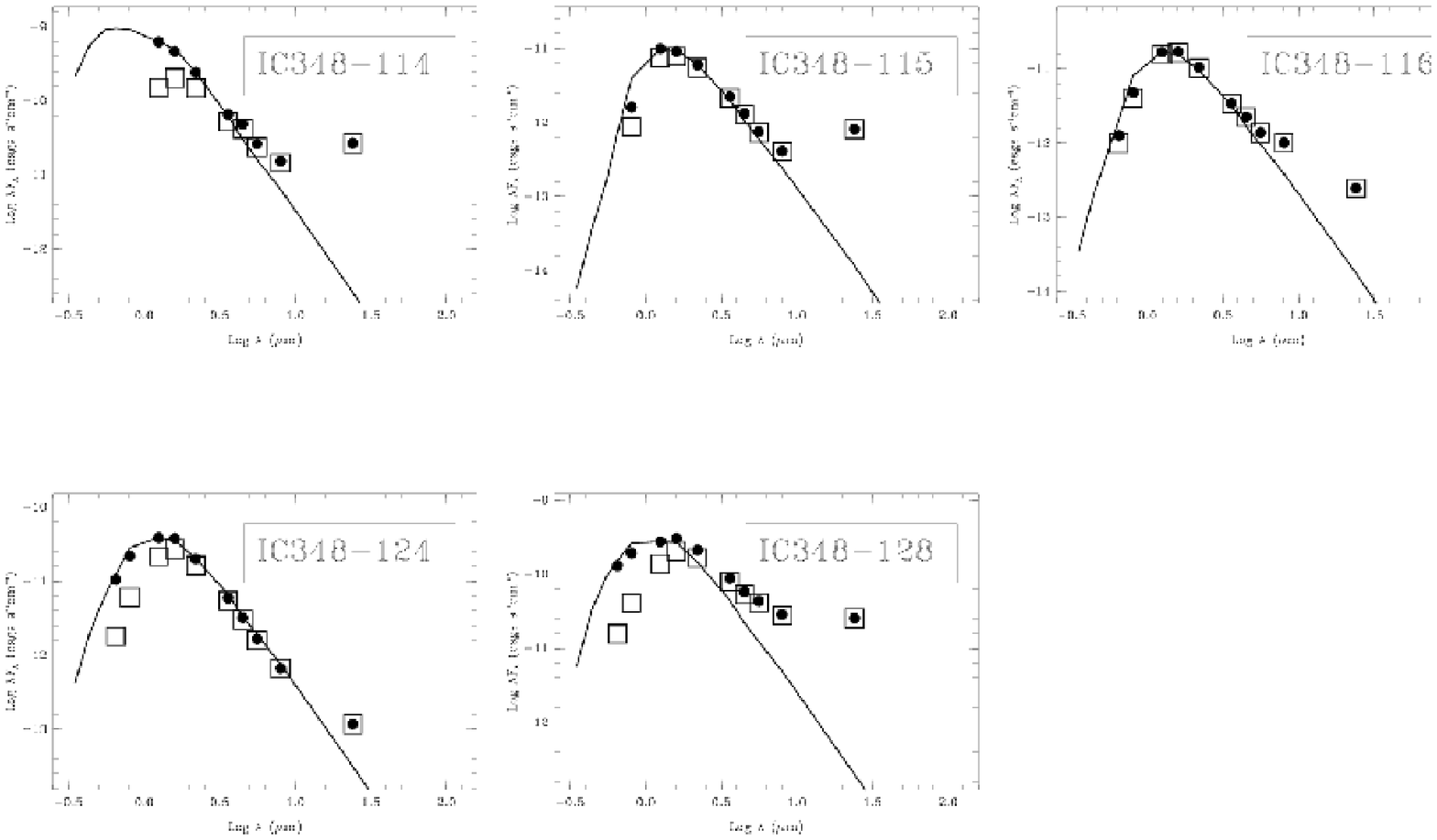}}\\[5mm]
\centerline{Fig. 4. --- Continued.}
\clearpage
\epsscale{1}
\begin{figure}
\figurenum{5a and 5b}
\scalebox{.44}{\includegraphics{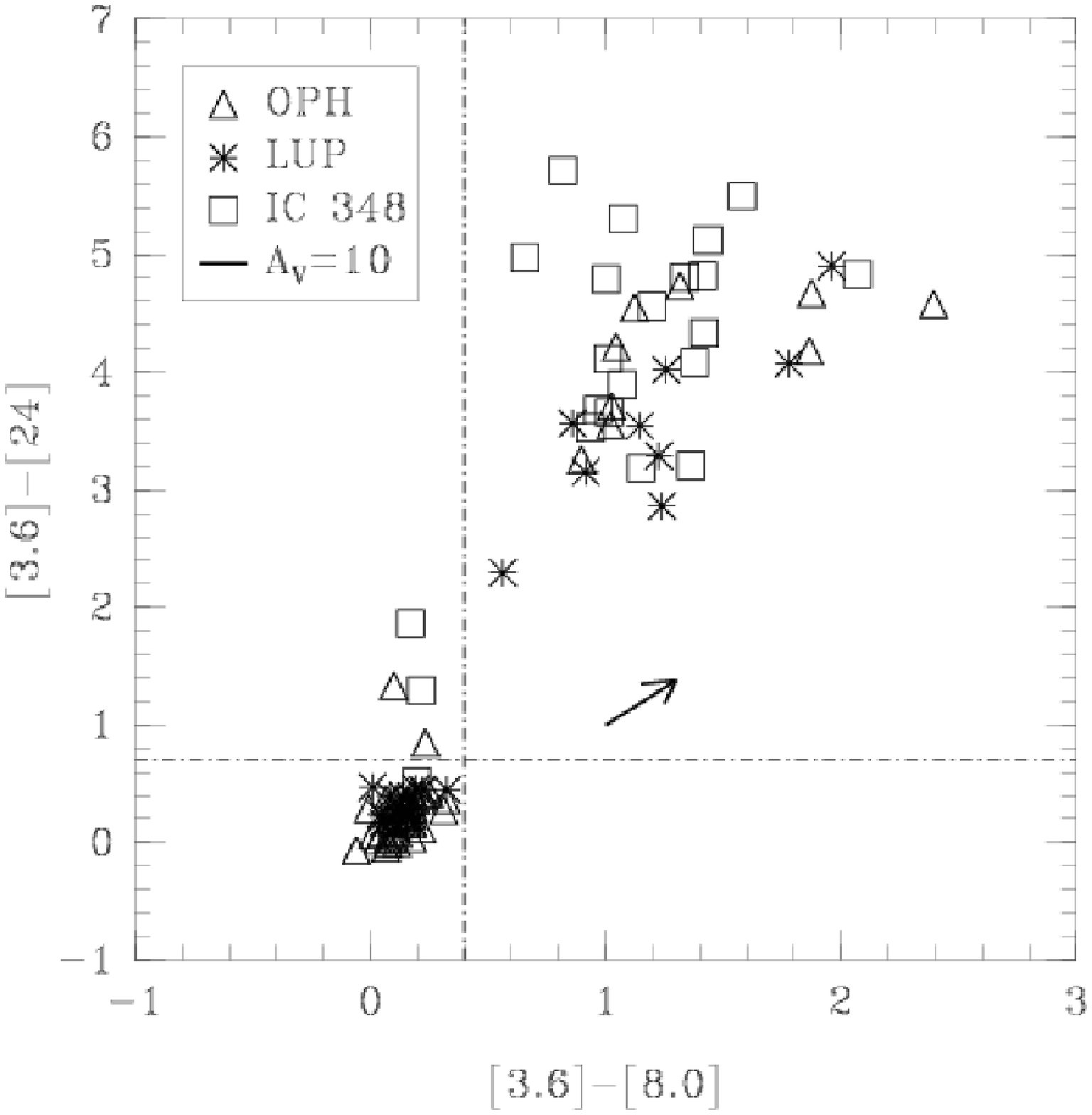}}
\scalebox{.45}{\includegraphics{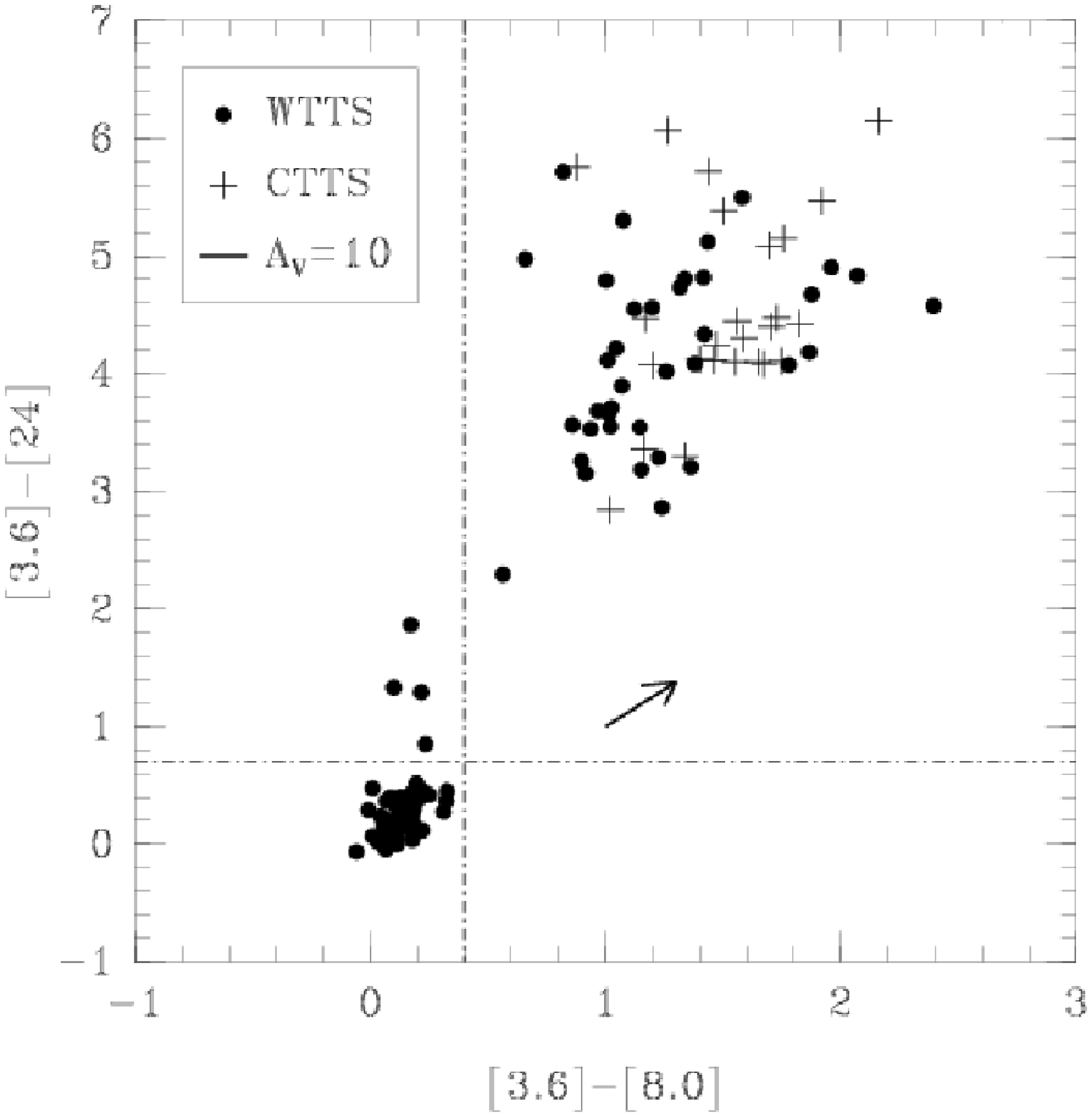}}
\caption{The Figure on the left shows the [3.6]--[24] vs. [3.6]--[8.0] colors of our sample of wTTs 
stars. Based on this diagram, we identify three different groups: (1) stellar photosphere with 
[3.6]--[24] $<$ 0.7 and  [3.6]--[8.0] $<$ 0.4, (2) Objects with [3.6]--[24] $>$ 0.7 and 
[3.6]--[8.0] $<$ 0.4 which show significant 24 $\mu$m excess but no evidence for 8.0 $\mu$m excess, and (3) 
objects with  [3.6]--[24] $>$ 0.7 and [3.6]--[8.0] $>$ 0.4 which show evidence for both IRAC and MIPS 
excesses. Objects in the second group are likely to have optically thin disks (see Section ~\ref{FDL}).
The Figure on the right combines our sample of wTTs with a sample of cTTs from 
Hartmann et al. (2005) and Lada et al. (2006).} 
 \end{figure}

\begin{figure}
\epsscale{1}
\figurenum{6a and 6b}
\plottwo{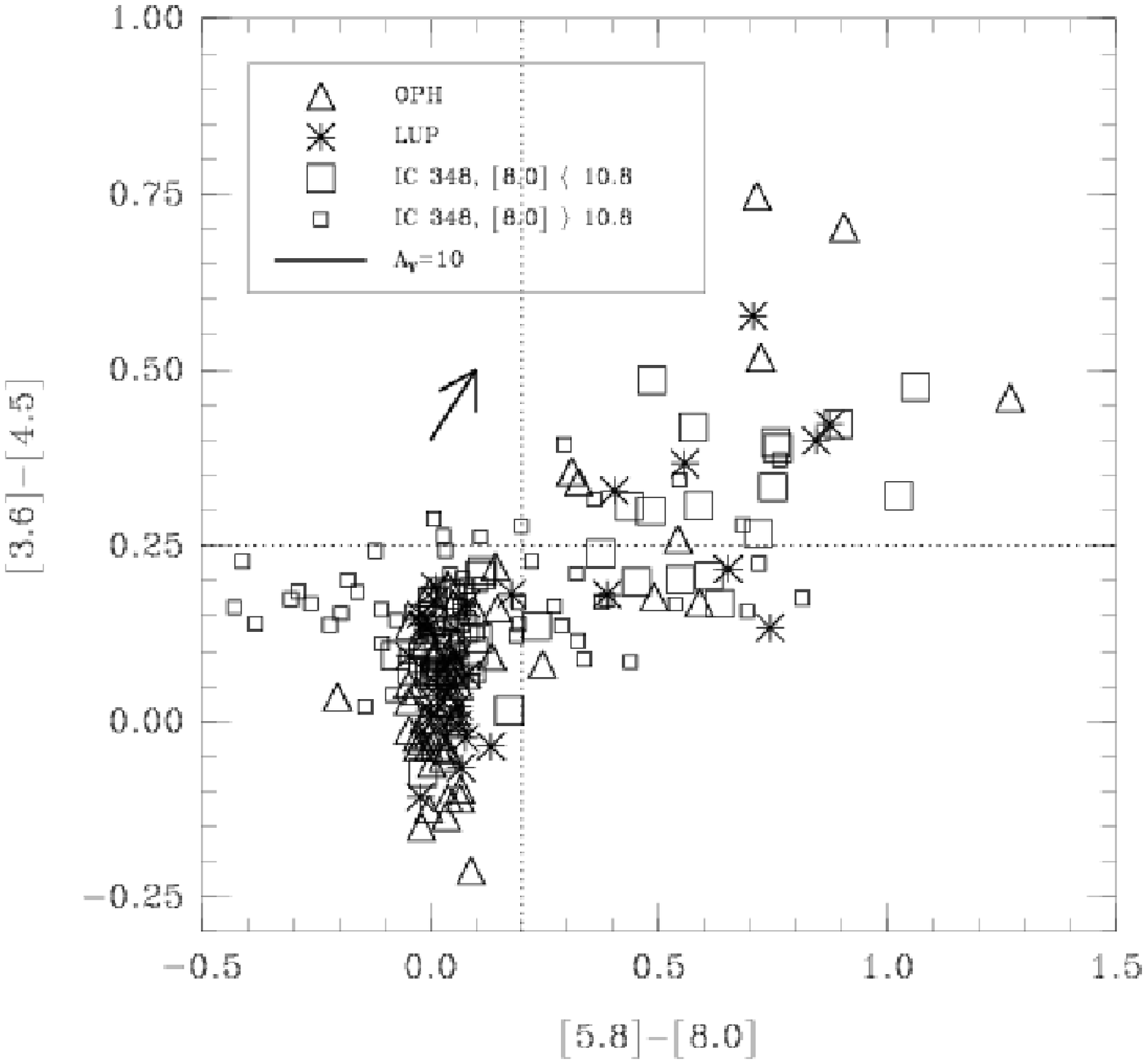}{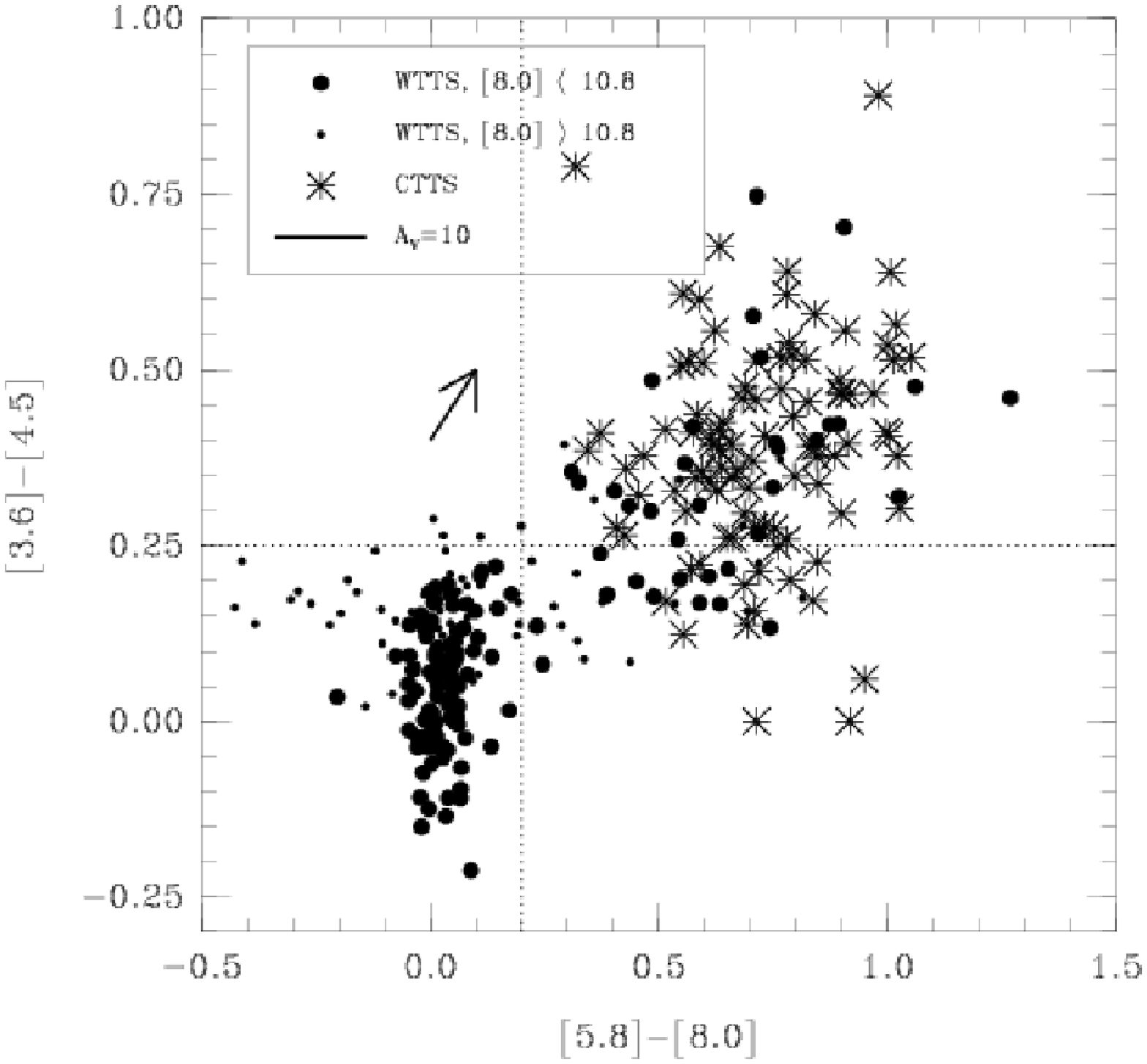}
\caption{The Figure in the left shows the [3.6]--[4.5] vs. [5.8]--[8.0] colors of our sample of wTTs. 
Faint IC 348 members ([8.0] $>$ 10.8 mag, 3 mJy) tend to have more uncertain colors than the rest of the sample 
and are shown as smaller open boxes. The dotted lines represent the 
approximate boundaries of the color of the stellar photospheres. Stars in the upper right corner of the 
diagram have both, 4.5 and 8.0 $\mu$m, excesses. Stars in the lower right corner of the diagram are 
stars with 8.0 $\mu$m excess but no 4.5 $\mu$m excess. The Figure in the right combines our sample of \
wTTs with cTTs from Hartmann et al. (2005) and Lada et al. (2006).}
\end{figure}

\begin{figure}
\figurenum{7a and 7b}
\epsscale{1}
\plottwo{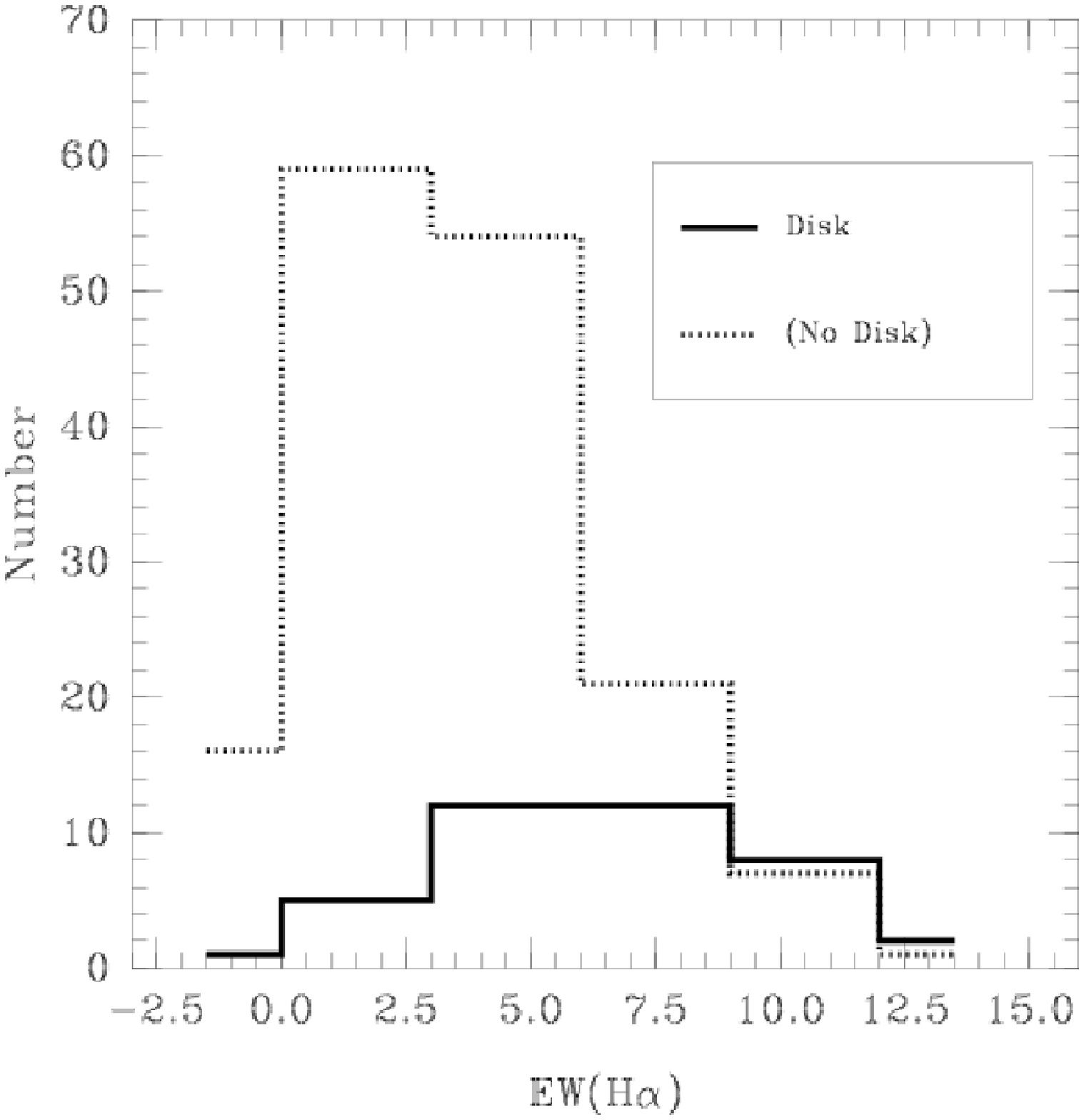}{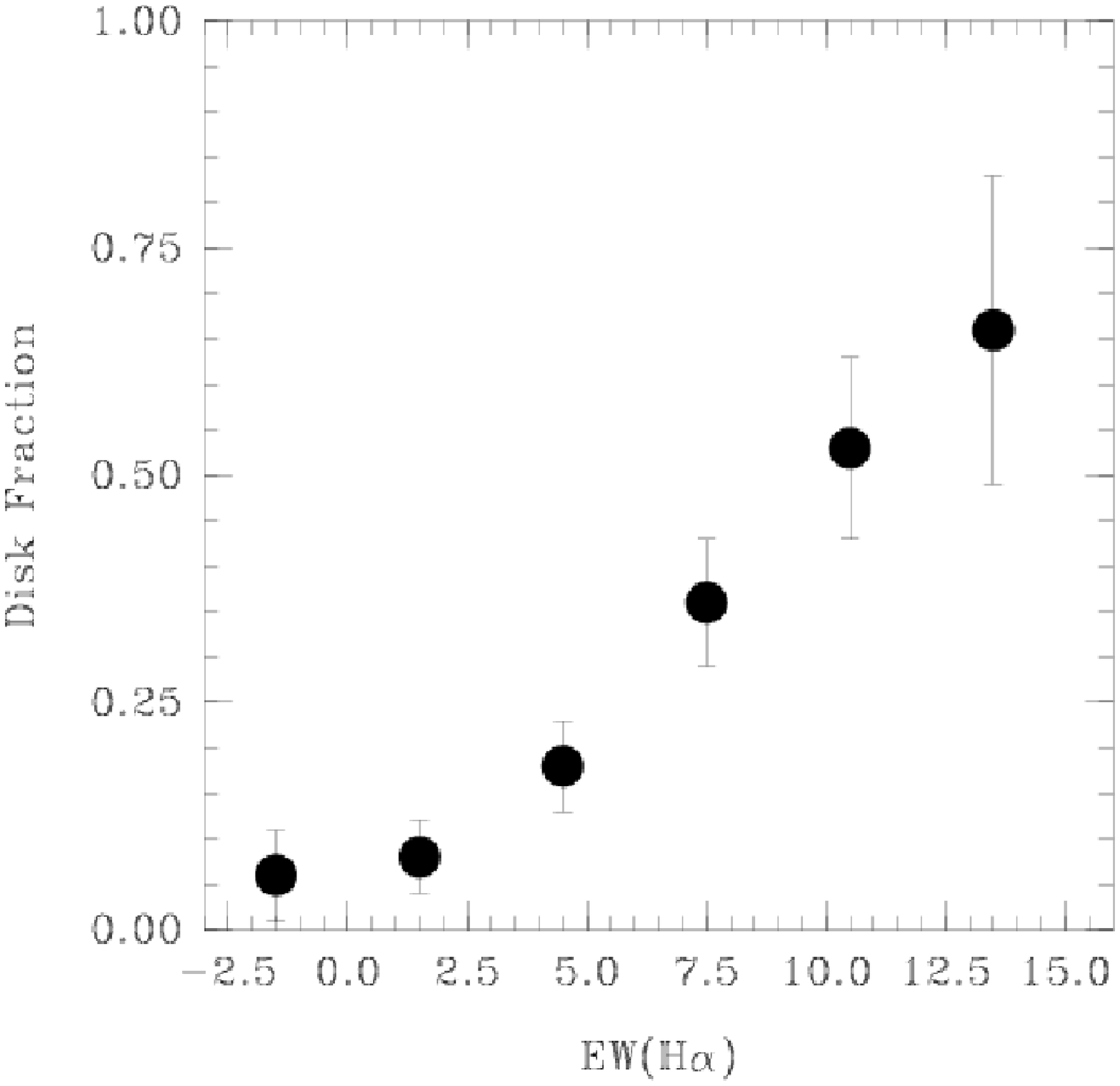}
\caption{Histogram of the $H\alpha$ equivalent width for stars with and without a disk (a),
and the disk fraction of wTTs vs. $H\alpha$ equivalent width (b). The disk fraction of wTTs 
seems to be a smooth function of $H\alpha$ equivalent width.}
\end{figure}

\begin{figure}
\figurenum{8}
\epsscale{1}
\plotone{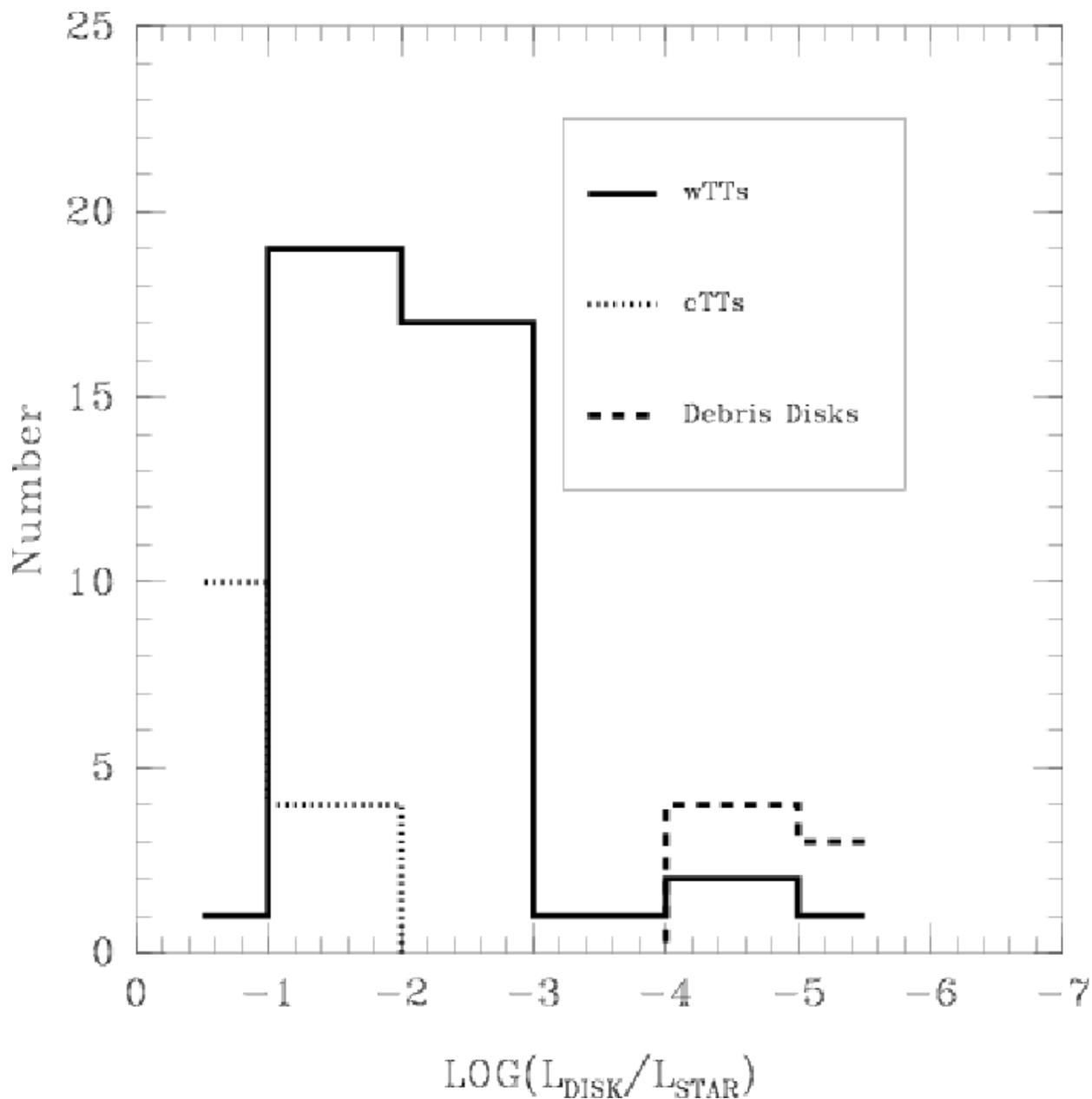}
\caption{The fractional disk luminosities, $L_{DISK}/L_{*}$, derived for our sample of wTTs disks.
The  $L_{DISK}/L_{*}$ values calculated using the same procedure for a sample of cTTs (From Cieza et al. 2005) 
and debris disks (from Chen et al. 2005) are shown for comparison.
The $L_{DISK}/L_{*}$ values of of wTTs disks fill the gap between the ranges observed for typical 
cTTs and debris disks, which are shown for comparison.}
\end{figure}

\begin{figure}
   \figurenum{9}
   \includegraphics[width=15cm,angle=0]{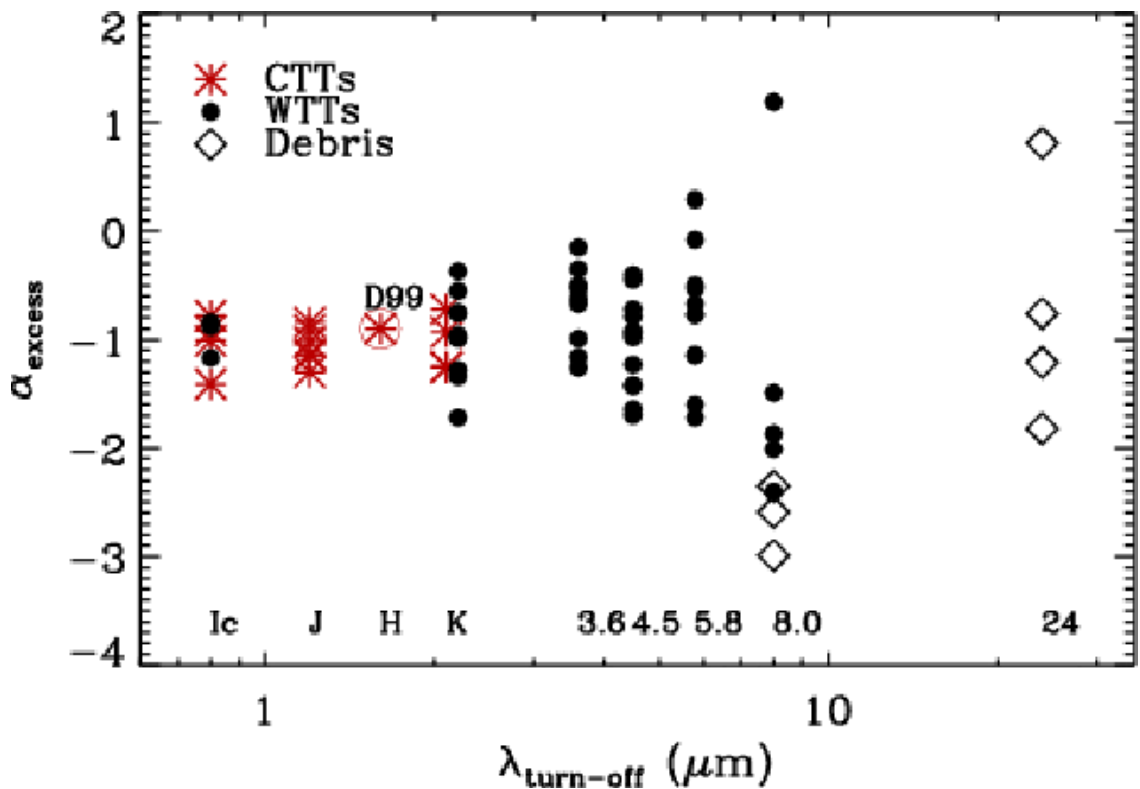}
   \caption{Distribution of excess slopes $\alpha_{\rm
   excess}$ vs. the wavelength at which the infrared
   excess begins $\lambda_{\rm turn-off}$ for the sample of wTTs
   (solid dots), a sample of cTTs in Chamaeleon from Cieza et al. (2005),
   the median SED of cTTs in Taurus from D'Alessio et al. (1999) in
   asterisks (marked as D99), and a sample of Debris disks from Chen
   et al. (2005) in diamonds. 
   The diagram shows a much larger spread in inner disk morphologies of wTTs with 
   respect of those of cTTs.}                                                                
\end{figure}

\begin{figure*}[tbp]
\centering
\figurenum{10}
\includegraphics[angle=90,width=\columnwidth,origin=bl]{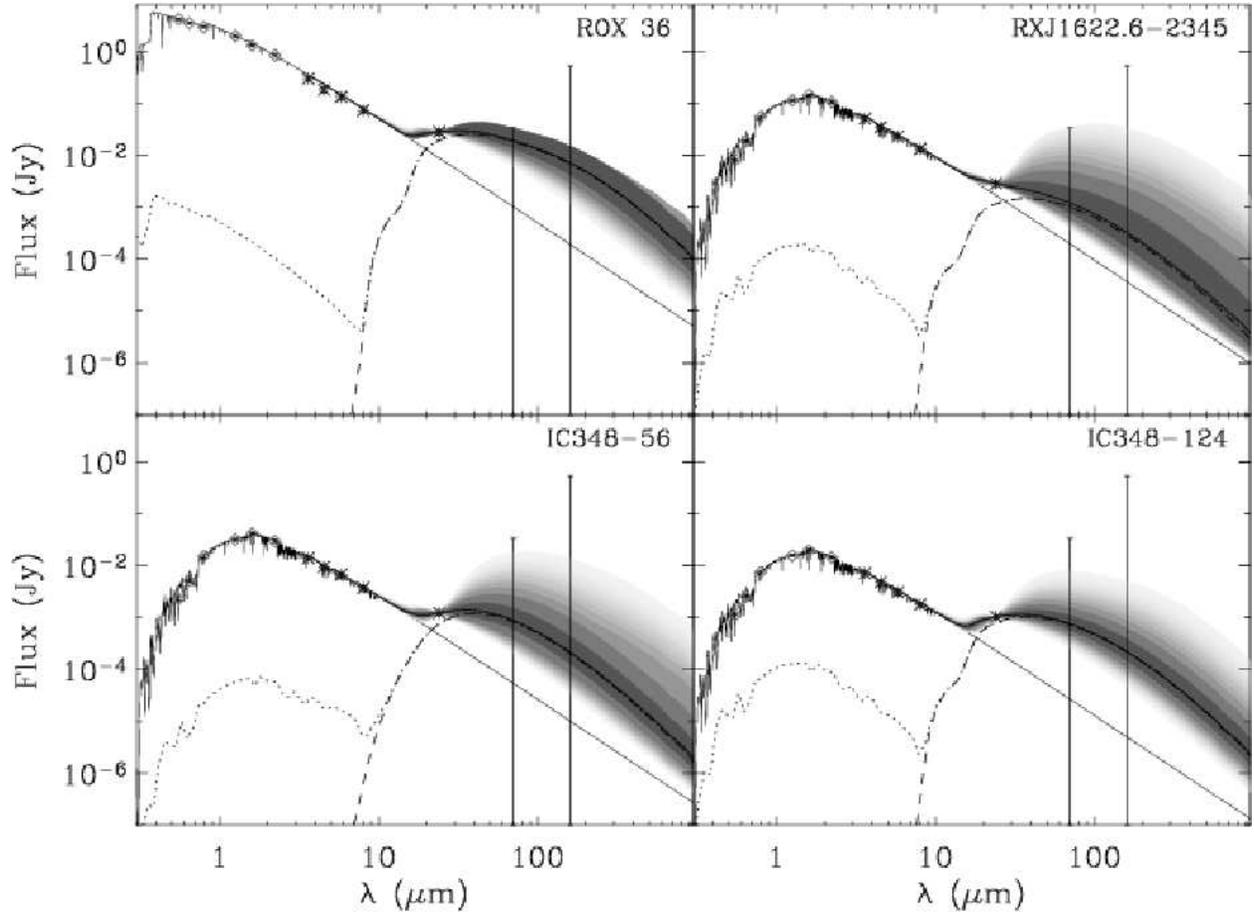}
\caption{wTTs with $24\,\mu$m excess consistent with optically thin
disks. On each plot, the darkest regions correspond to the most likely fits 
to the SEDs. The dashed line shows the thermal emission for the best-fit model, 
while the dotted line corresponds to the total disk emission (i.e. including scattered light emission.)}
\label{JC_dds_SEDs}
\end{figure*}

\begin{figure}[tbp]
\centering
\epsscale{0.8}
\figurenum{11}
\hbox to \textwidth
{
\parbox{0.33\textwidth}{
\includegraphics[angle=0,width=0.35\columnwidth,origin=bl]{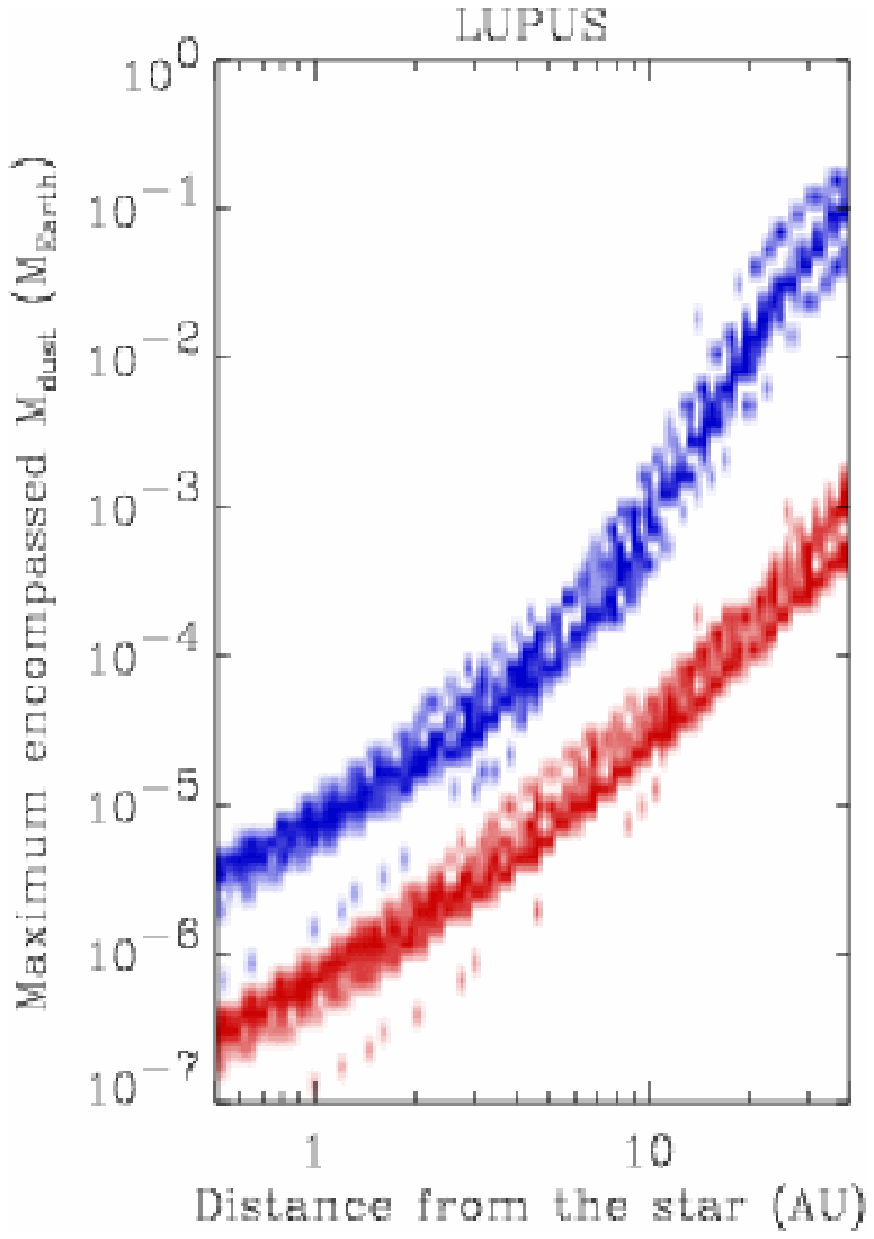}
}
\hspace{0.4cm}
\parbox{0.33\textwidth}{
\includegraphics[angle=0,width=0.35\columnwidth,origin=bl]{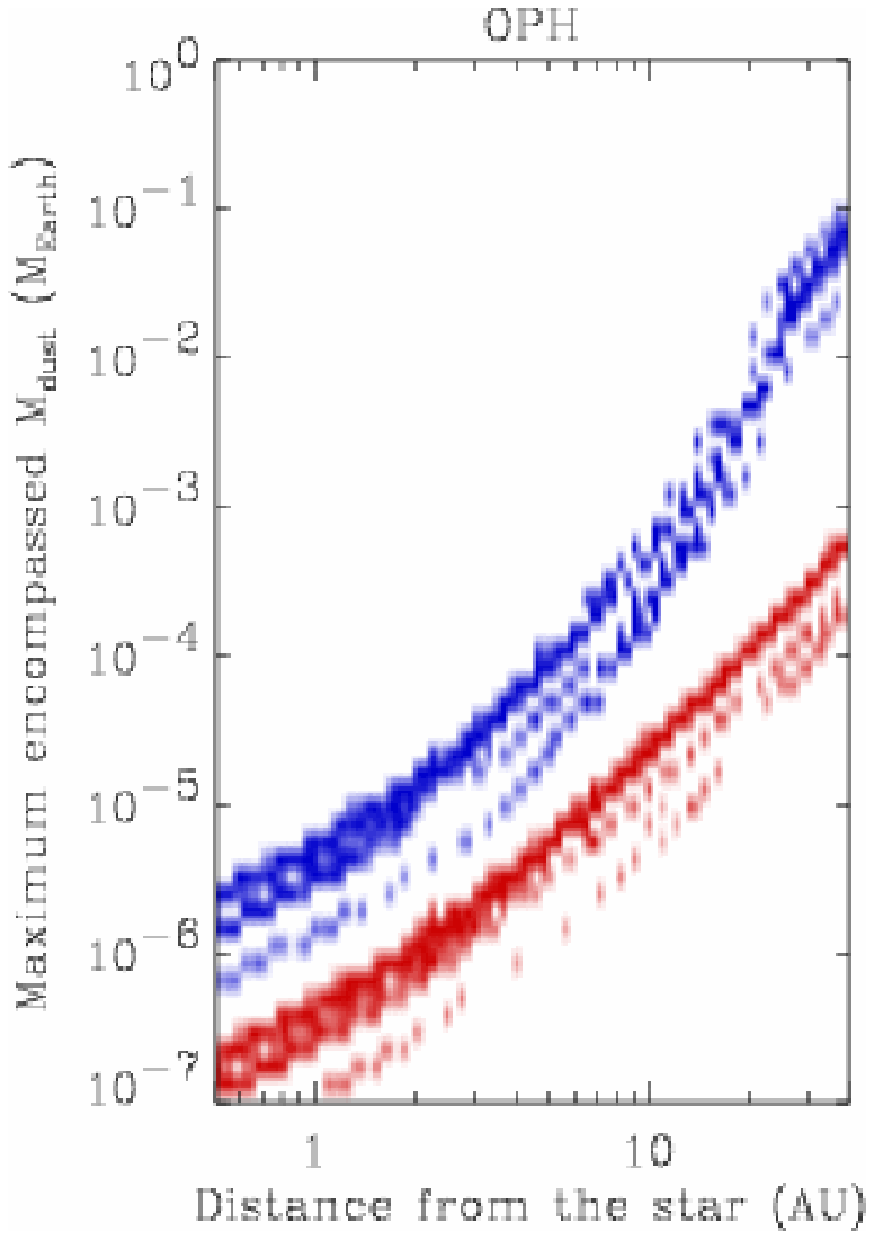}
}
}
\caption{Maximum encompassed dust mass as a function of the distance from the star for
the Lupus and Ophiuchus  clouds (respectively left and right panels).
The red area corresponds to mass upper limits when minimum grain sizes $a_{\rm min}$
between $0.05\,\mu$m and $0.5\,\mu$m are considered, while the blue area corresponds
to $10\,   \mu$m$   <  a_{\rm min}  < 100\,   \mu$m.
}
\label{MvsR}
\end{figure}

\begin{figure}
\figurenum{12a and 12b}
\plottwo{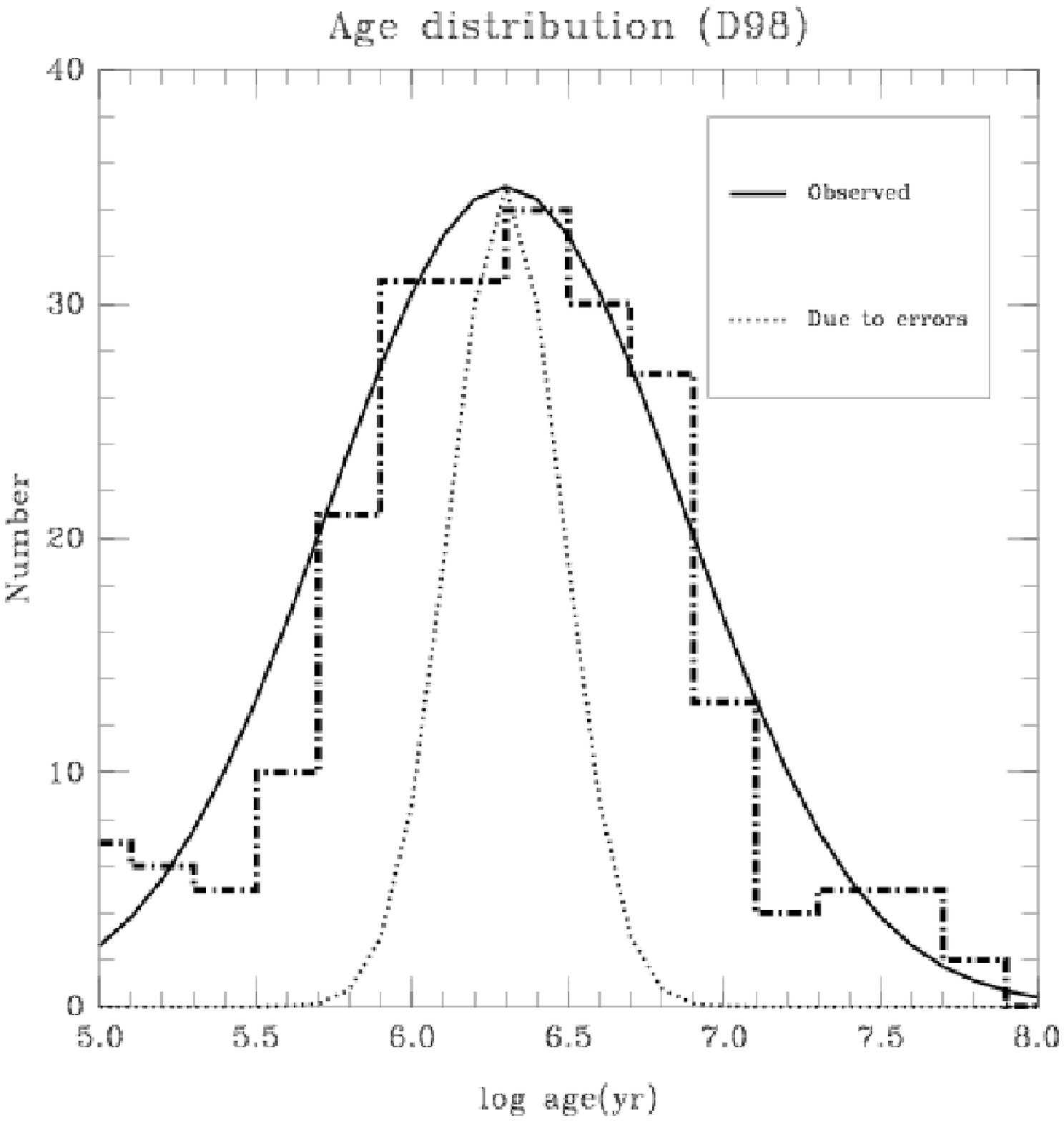}{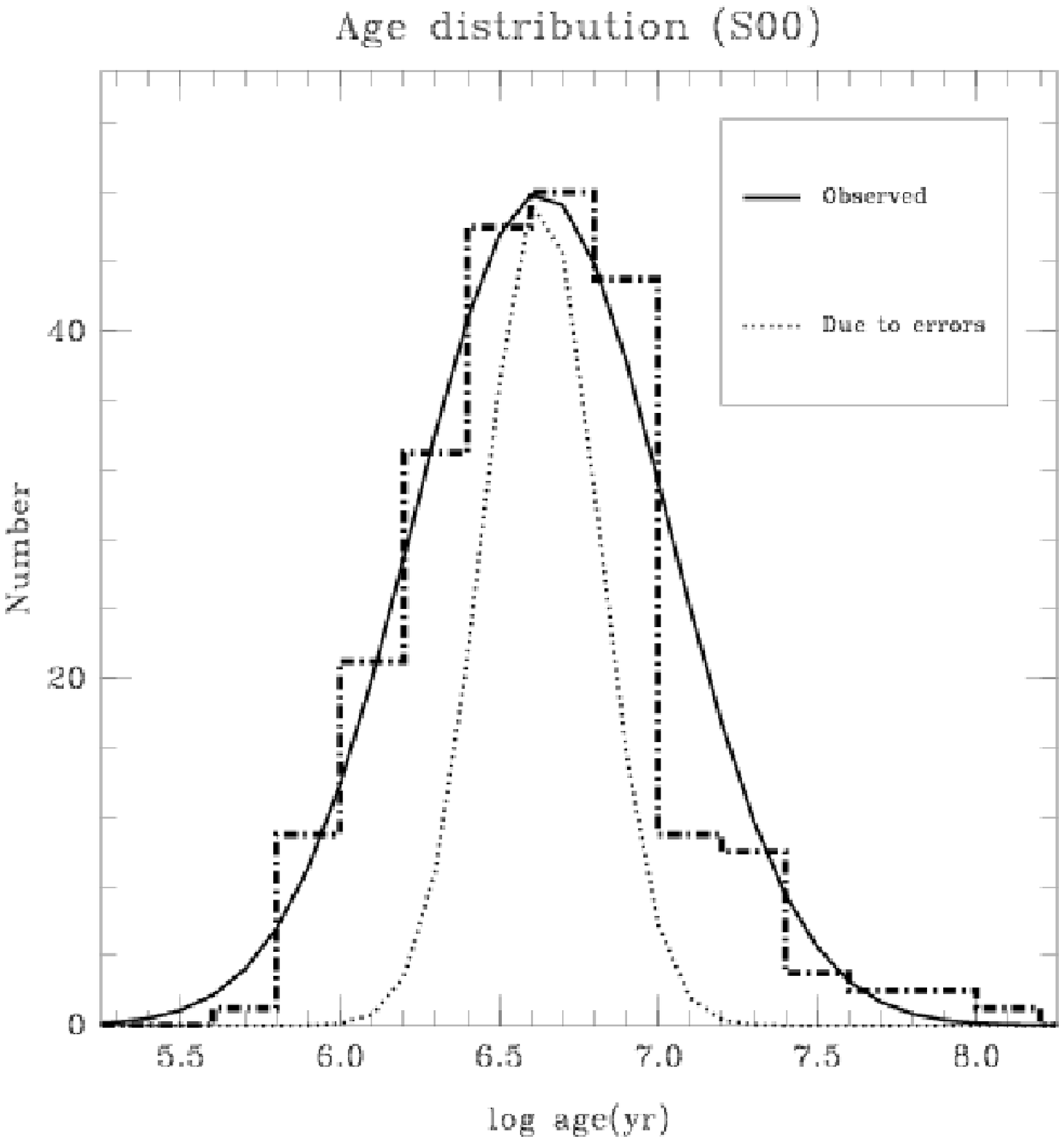}
\caption{The distributions of stellar ages obtained for our sample using the D98 and S00 evolutionary tracks 
(left and right panels, respectively). The observed age distributions (solid lines) are significantly 
wider than what is expected from the propagation of observational errors into derived stellar ages 
(dotted lines).}
\end{figure}

\begin{figure}
\figurenum{13a and 13b}
\plottwo{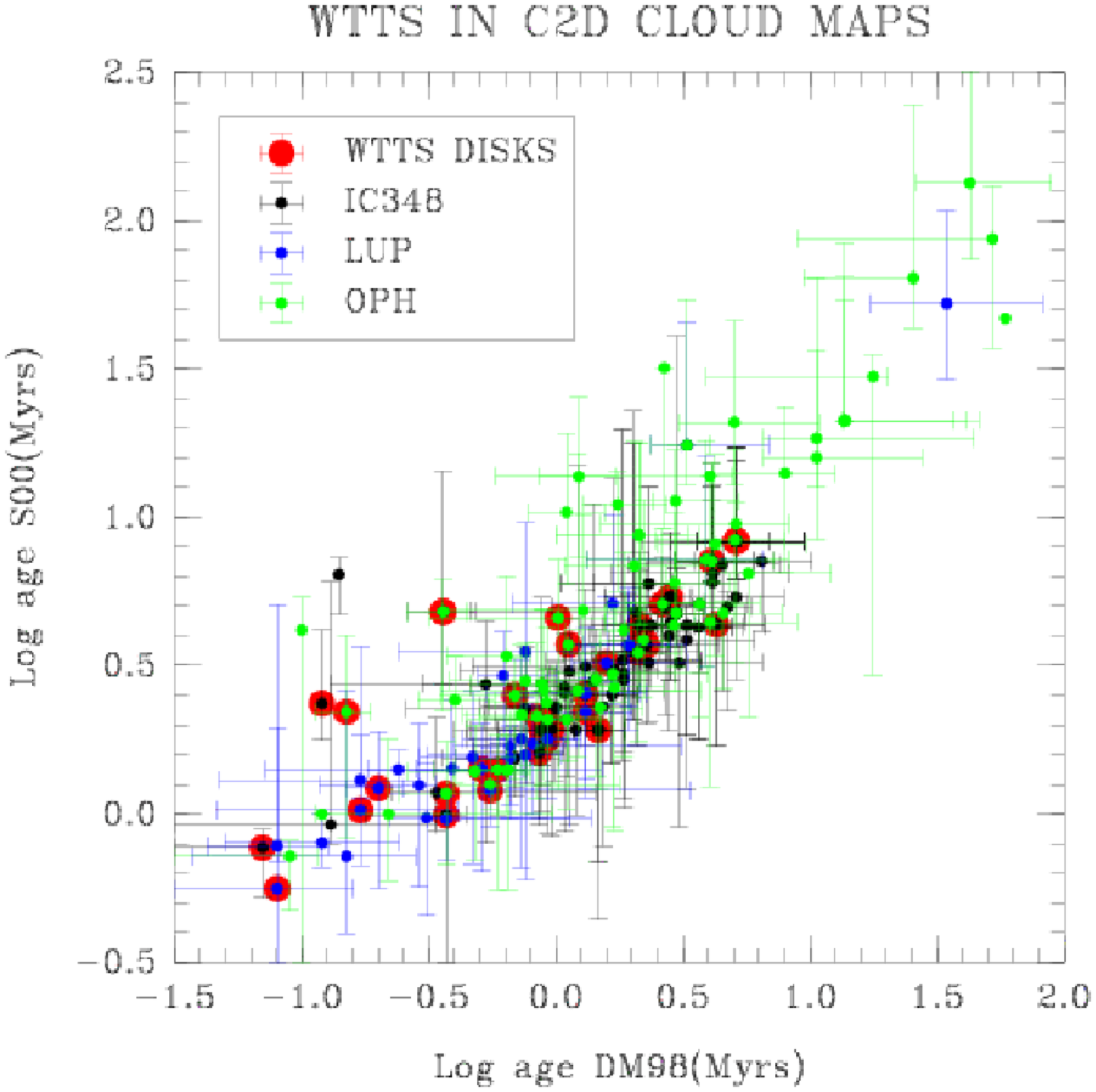}{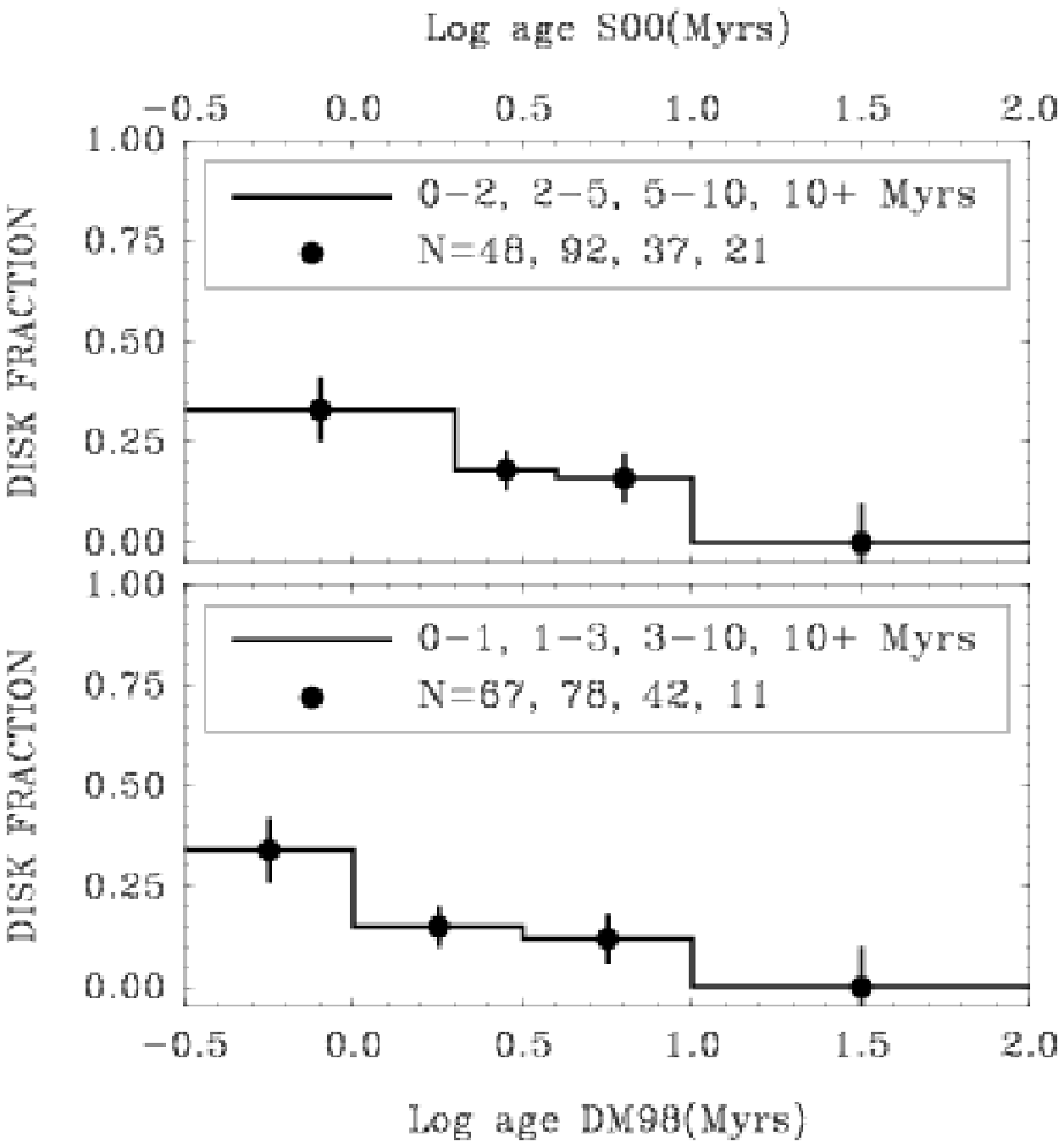}
\caption{The stellar ages derived for our sample of wTTs using two different sets of evolutionary tracks 
(D98 and S00). The error bars have been calculated adopting a $T_{eff}$ uncertainty equal to one spectral 
type subclass and a luminosity error calculated from the uncertainty in the distance (a). A clear decrease in the disk 
fraction is seen with increasing age. $\sim$40$\%$ of the targets that are both younger than 1 Myrs according 
to D98 tracks and younger than 2 Myrs according to S00 tracks have disks. None of the stars that are older than 
$\sim$10 Myrs according to either of the models have disks (b).}
\end{figure}

\clearpage

\end{document}